\documentclass[aps,prb,superscriptaddress,twocolumn]{revtex4-2}

\usepackage{setspace}
\usepackage{lmodern}

\usepackage[latin9]{inputenc}
\usepackage{amsmath}
\usepackage{amssymb}
\usepackage{graphicx}
\usepackage{epstopdf}
\usepackage{color}
\usepackage{enumerate}
\usepackage{ulem}
\usepackage{bm} 
\usepackage{physics}

\usepackage{array}
\usepackage[labelfont=bf]{caption}
\usepackage{multirow}
\usepackage{floatpag}
\usepackage[misc]{ifsym}
\usepackage{sistyle}
\usepackage{upgreek}

\begin{document}

\title{Evidence for correlated electron pairs and triplets in quantum Hall interferometers}

\author{Wenmin Yang}
\altaffiliation{These authors contributed equally to this work.}
\affiliation{Univ. Grenoble Alpes, CNRS, Grenoble INP, Institut N\'{e}el, 38000 Grenoble, France}
\author{David Perconte}
\altaffiliation{These authors contributed equally to this work.}
\affiliation{Univ. Grenoble Alpes, CNRS, Grenoble INP, Institut N\'{e}el, 38000 Grenoble, France}
\author{Corentin D\'{e}prez}
\affiliation{Univ. Grenoble Alpes, CNRS, Grenoble INP, Institut N\'{e}el, 38000 Grenoble, France}
\author{Kenji Watanabe}
\affiliation{Research Center for Electronic and Optical Materials, National Institute for Materials Science, 1-1 Namiki, Tsukuba 305-0044, Japan}
\author{Takashi Taniguchi}
\affiliation{Research Center for Materials Nanoarchitectonics, National Institute for Materials Science,  1-1 Namiki, Tsukuba 305-0044, Japan}
\author{Sylvain Dumont}
\affiliation{Univ. Grenoble Alpes, CNRS, Grenoble INP, Institut N\'{e}el, 38000 Grenoble, France}
\author{Edouard Wagner}
\affiliation{Univ. Grenoble Alpes, CNRS, Grenoble INP, Institut N\'{e}el, 38000 Grenoble, France}
\author{Fr\'{e}d\'{e}ric Gay}
\affiliation{Univ. Grenoble Alpes, CNRS, Grenoble INP, Institut N\'{e}el, 38000 Grenoble, France}
\author{In\`{e}s Safi}
\affiliation{Universit\'{e} Paris-Saclay, CNRS, Laboratoire de Physique des Solides, 91405 Orsay, France}
\author{Hermann Sellier}
\affiliation{Univ. Grenoble Alpes, CNRS, Grenoble INP, Institut N\'{e}el, 38000 Grenoble, France}
\author{Benjamin Sac\'{e}p\'{e}}
\email{benjamin.sacepe@neel.cnrs.fr}
\affiliation{Univ. Grenoble Alpes, CNRS, Grenoble INP, Institut N\'{e}el, 38000 Grenoble, France}

\begin{abstract}
\bf{The pairing of electrons is ubiquitous in electronic systems featuring attractive inter-electron interactions, as exemplified in superconductors\cite{TinkhamBook2004}. Counter-intuitively, it can also be mediated in certain circumstances by the repulsive Coulomb interaction alone\cite{Hamo2016, Hong2018}. Quantum Hall (QH) Fabry-P\'{e}rot interferometers (FPIs) tailored in two-dimensional electron gas under a perpendicular magnetic field have been argued to exhibit such unusual electron pairing seemingly without attractive interactions \cite{choi2015,sivan2018,nakamura2019, Biswas2023}. 
Here, we show evidence in graphene QH FPIs\cite{deprez2021,ronen2021,zhao2022,Fu23} revealing not only a similar electron pairing at bulk filling factor $\nu_{\rm{B}}=2$ but also an unforeseen emergence of electron tripling characterized by a fractional Aharonov-Bohm (AB) flux period $h/3e$ ($h$ is the Planck constant and $e$ the electron charge) at $\nu_{\rm{B}}=3$. 
Leveraging novel plunger-gate spectroscopy, we demonstrate that electron pairing (tripling) involves correlated charge transport on two (three) entangled QH edge channels. This spectroscopy indicates a quantum interference flux-periodicity determined by the sum of the phases acquired by the distinct QH edge channels having slightly different interfering areas. Phase jumps observed in the pajama maps can be accounted for by the frequency beating between pairing/tripling modes and the outer interfering edge.
Though our discovery of three entangled QH edge channels with apparent electron tripling defies understanding, it provides an important selection constraint for any theoretical explanation. In addition, this phenomenon differs from superconducting pairs and highlights the role of isolated edges as well as inter-edge interactions.}
\end{abstract}

\maketitle 

The quantum Hall effect is known to host a wide range of correlated and symmetry-protected phases. Coulomb repulsion plays a central role in it, shaping the structure of QH edge channels\cite{Chklovskii92}, inducing (pseudo) spin-polarized QH ferromagnets\cite{Ezawa2009}, or generating fractional quantum Hall states\cite{halperin2020} with anyonic excitations that may be useful for topological quantum computation\cite{nayak2008}. 

In 2015, a surprise came with the observation of the pairing of electrons in QH interferometers. Choi and co-workers\cite{choi2015} found in GaAs Fabry-P\'{e}rot interferometers (FPIs) defined by two quantum point contacts (QCPs) in series\cite{van1989} an anomalous Aharonov-Bohm (AB) effect with halved flux-periodicity, $h/2e$. The specific configurations identified were the presence of at least two QH edge channels in the FPI, that is, a bulk filling factor $\nu_{\rm{B}} > 2$, and interference from the outer channel while the inner one forms a closed loop. Strikingly, this electron pairing was confirmed by quantum shot noise that evidenced an effective charge $e^*\sim 2e$ (Refs.\cite{choi2015,Biswas2023}), pointing conspicuously towards correlated electron-pair transport.
	
The analogy with Cooper pairing in superconductors is tantalizing, however, the resemblance is only apparent since there are no attractive, non-Coulombic interactions, nor evidence of a macroscopic condensate. On the theoretical front, most efforts to date have failed to describe this baffling phenomenon\cite{Ferraro2017,frigeri2019,Frigeri2020}. Yet, an effective dynamical pairing via the exchange of neutralons\cite{Frigeri2020} has been put forward, but cannot capture all phenomenology\cite{choi2015,sivan2018,Biswas2023}. 
	
Here, we opt for a different platform --the graphene QH FPI \cite{deprez2021,ronen2021,zhao2022,Fu23} to uncover new insights into this phenomenon. By leveraging the high-tunability of its plunger gate\cite{deprez2021} and conducting systematic out-of-equilibrium transport measurements, we establish a new QH edge channel spectrometry that allows us to identify the exact channels involved coherently in the electron pairing. This spectroscopy allows us to conclude that pairing consistently occurs when the number of edge channels exceeds one. The pairing's weight significantly grows with an increase in the filling factor, ultimately leading to visible frequency beating in the pajama map at a filling factor higher than 2.5. Furthermore, at filling factor $\nu_{\rm{B}}=3$, we uncover evidence of correlated transport involving three electrons over the three distinguishable edge channels. Our unprecedented systematic exploration of the flux and energy bias parameter space gives key insights into a complex interplay between edge channels and their interactions.	

\begin{figure*}[ht]
\includegraphics[width=12cm]{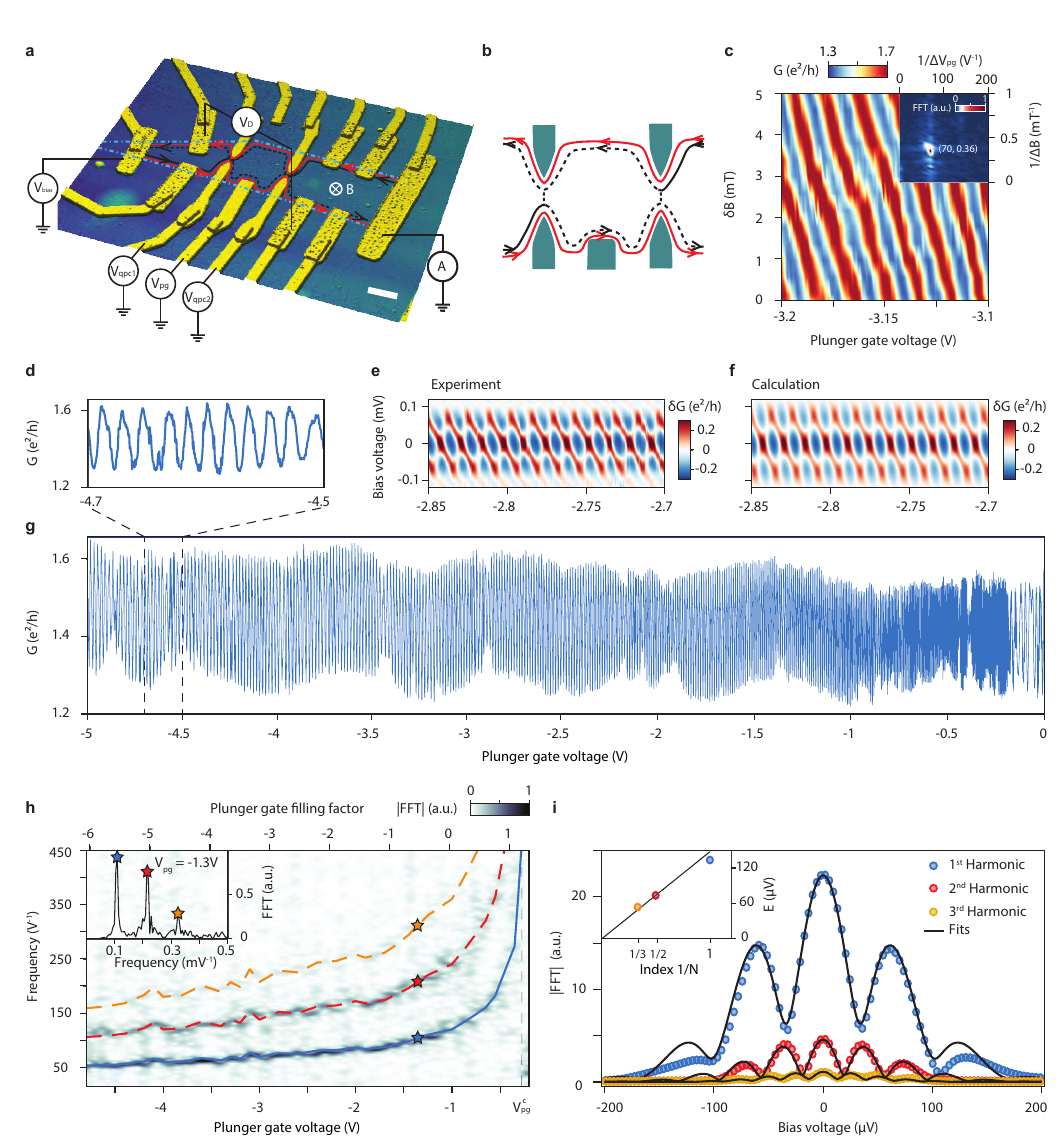}
\centering
\caption{\textbf{a,} Atomic force microscope image of the hBN encapsulated graphene heterostructure. Ohmic contacts are highlighted in yellow (rough surface), while QPC electrodes and the plunger gate are in light yellow (smooth surface). Inner and outer QH edge channels at $\nu_{\rm{B}} = 2$ are depicted in black and red, respectively. The dashed blue line represents the graphene, as it is not visible here.
Here the inner edge channel is partially transmitted by the QPCs. The measured diagonal voltage $V_D$ and current $I$ give the diagonal conductance. 
\textbf{c,} Pajama map obtained by partitioning the inner channel at $V_{\rm{bg}} = 1.8~\rm{V}$ ($\nu_{\rm{B}} = 2.26$) with QPC transmissions of $T_{\rm{1}} = 0.67$ and $T_{\rm{2}} = 0.66$, as schematically shown in panel (\textbf{b}). The negative slope of the constant phase lines indicates AB-dominated interference \cite{Zhang2009}. Inset is the FFT of the pajama map. \textbf{g,} Diagonal conductance oscillations across a wide range of plunger-gate voltages, using the same configuration as in panel \textbf{c}. The oscillations shown here are plotted at zero bias, while their dependence on bias voltage is depicted in Supplementary Video 1.
\textbf{d,} Focus on oscillations within a narrow $V_{\rm{pg}}$ range between -4.7~\rm{V} and -4.5~\rm{V}. 
\textbf{e,} Checkerboard pattern extracted from bias voltage dependent oscillation in Supplementary Video 1. 
\textbf{f,} Calculated checkerboard pattern in the presence of three harmonics (see Methods). 
\textbf{h,} Sliding FFT is applied to analyze the oscillations shown in panel \textbf{g}. The FFT is computed over a window of $0.175~\rm{V}$. This gate spectroscopy is obtained at $V_{\rm{DC}} = -28~\upmu\rm{V}$, and its full energy dependence can be seen in Supplementary Video 2. The blue solid line emphasizes the first harmonic oscillation. The dashed lines in red and yellow represent the calculations for twice and three times the frequency of the blue line, respectively. The inset displays the FFT at a fixed $V_{\rm{pg}} = -1.3~\rm{V}$, revealing three harmonic peaks marked with blue, red, and yellow stars. 
\textbf{i,} The lobe structure of the three Fourier harmonics allows for the extraction of the Thouless energy of each harmonic (see Methods). Inset: The Thouless energy shows a nearly linear relationship with the inverse of the harmonic index.}
\label{Fig1}
\end{figure*} 

\begin{figure*}[ht]
\includegraphics[width=16cm]{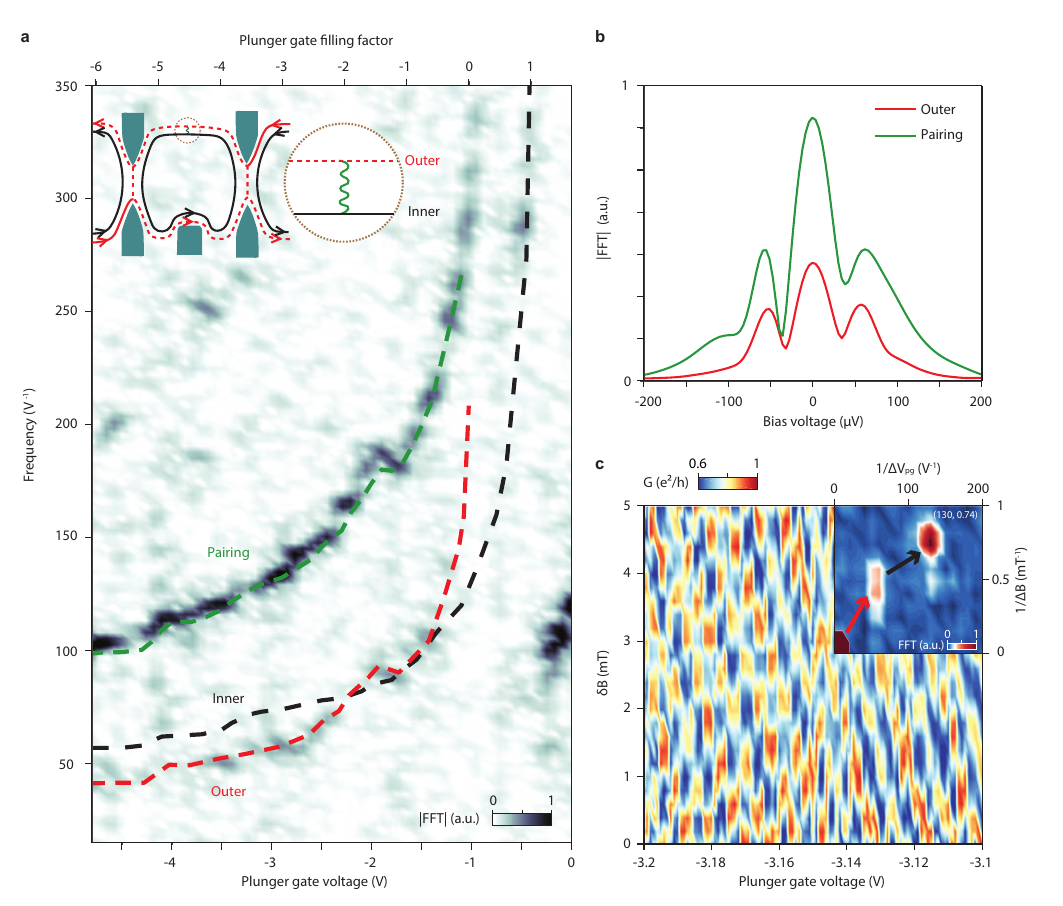}
\centering
\caption{\textbf{Gate-spectroscopy of the outer edge channel interference.} 
Conductance oscillation was measured by partitioning the outer edge channel with QPC transmission of $T_{\rm{1}} = 0.91 $ and $T_{\rm{2}} = 0.94 $ at $V_{\rm{bg}} = 1.8~\rm{V}$ ($\nu_{\rm{B}} = 2.26$). \textbf{a}, Fourier transform amplitude of the oscillation is plotted versus plunger-gate voltage and plunger-gate oscillation frequency at $V_{\rm{DC}} = -55.5~\upmu \rm{V}$, the Fourier transform is calculated over a $0.2~\rm{V}$ window which slides over the whole plunger gate voltage range. Supplementary Video 3 shows the dependency of this frequency dispersion on DC bias voltage. The black and red dashed lines represent the frequency dispersion of the inner and outer edges, respectively. The green dashed line is the sum of these two frequencies and coincides with the strong signal attributed to pairing. The edge channel configuration is shown in the inset schematics, where the red dashed line indicates the partitioned outer channel and the green wavy line represents the inter-channel interaction. \textbf{b}, The lobe structures of the pairing (green-colored) and outer (red-colored) channels were extracted from the bias-voltage dependent frequency dispersion in Supplementary Video 3. 
\textbf{c}, The pajama map shows frequency beating between the outer channel and pairing at zero bias (Supplementary Video 4 provides the entire bias-voltage dependent pajama maps). Inset: The 2D Fourier transform of the pajama map reveals two frequencies originating from the outer channel and pairing. The pairing frequency is depicted as the sum of the outer channel frequency (red arrow) and the inner channel frequency determined from Fig. \ref{Fig1}h (black arrow).}
\label{Fig2}
\end{figure*}

The QH FPIs are made with hBN-encapsulated graphene deposited onto a graphite gate acting as a back-gate electrode. Two QPCs are electrostatically defined by a set of two palladium split-gate electrodes\cite{Zimmermann2017,deprez2021}. The FPIs are equipped with a plunger-gate electrode to tune the effective area enclosed by the QH edge channels. Several 1D ohmic contacts\cite{Wang2013} allow us to source and drain current and probe voltages across the FPI. Fig. \ref{Fig1}a shows an atomic force microscopy topography of the device studied in the main text, which has an FPI cavity area of $2.2 \pm 0.2\,\upmu$m$^2$. Importantly, the FPI is defined by the pristine, non-etched edges of the graphene flake, ensuring confinement of the QH edge channels to within a few magnetic lengths of the crystal edge, without any edge reconstruction\cite{coissard2023}, as well as by split and plunger gates. 	
All experiments are performed at a magnetic field of 14 T and a temperature of 0.01 K. Partial pinch{-}off of the inner channel as overlaid in Figs. \ref{Fig1}a-b yields conductance oscillations shown in Fig. \ref{Fig1}c with negative slope in the magnetic field, $B$, versus plunger gate voltage, $V_{\rm{pg}}$, plane, which is characteristic of AB quantum interference for a flux periodicity of $h/e$\cite{Zhang2009,deprez2021}.

\bigskip
\textbf{Gate-spectroscopy fingerprint of QH edge channels}
\bigskip

\begin{figure*}[!ht]
\includegraphics[width=16cm]{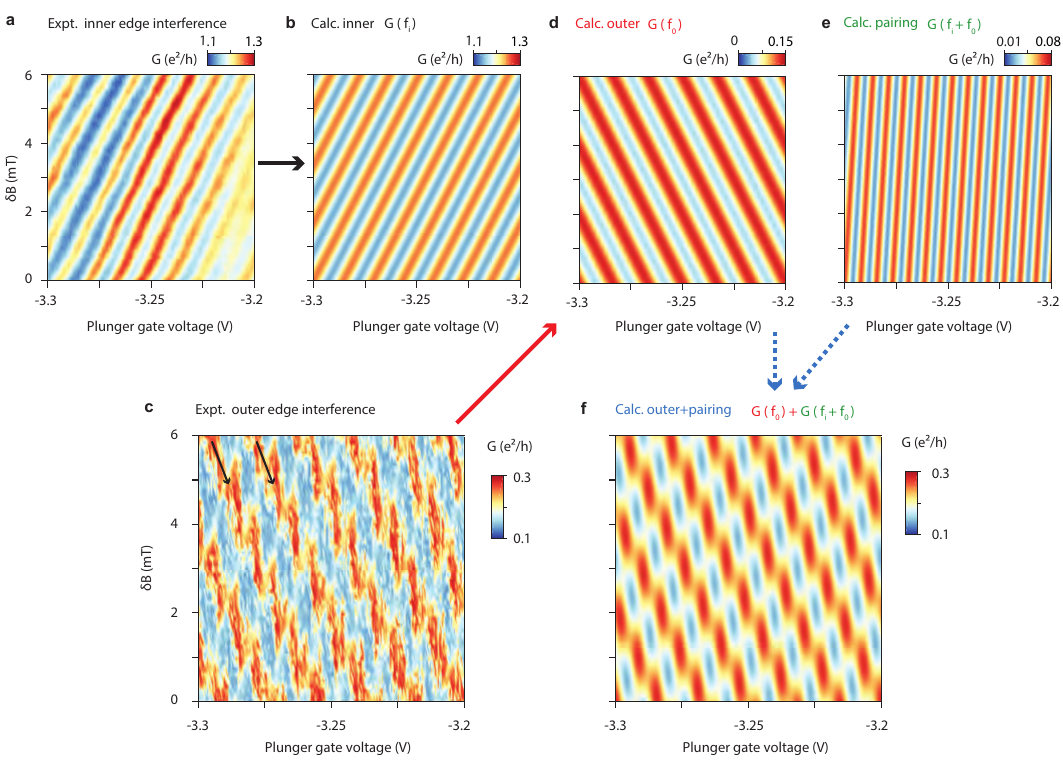}
\centering
\caption{\textbf{Experimental and simulated analysis of phase jumps in the pajama measured at $V_{\rm{bg}} = 1.2~\rm{V}$ ($\nu = 1.7$)}. \textbf{a,} The pajama map, obtained by partitioning the inner edge channel, suggests a single-frequency oscillation. \textbf{b,} Calculated pajama map with the oscillation frequency of the inner edge channel derived from the panel \textbf{a}. \textbf{c,} Observed pajama map when partitioning the outer edge channel, where the dominant frequency is associated with the bare outer edge. \textbf{d,} Calculated pajama map of the bare outer channel using the dominant frequency observed in panel \textbf{c}. \textbf{e,} The pajama map featuring the oscillation frequency of electron pairing, is calculated by summing up the frequencies of the inner and outer channels. \textbf{f,} The beating phenomenon in the pajama map occurs due to the simultaneous presence of two oscillations at the pairing and outer edge frequencies.}
	\label{Fig3}
\end{figure*}

The considerable advantage of graphene FPI over conventional semiconductors is the absence of a bandgap, which allows one to achieve a very large electrostatic tuning of the charge carrier density ranging from the electron states to the hole states. 
Fig. \ref{Fig1}g illustrates this tunability with conductance oscillations versus plunger-gate voltage, $V_{\rm{pg}}$, from -5 to 0 V, reflecting the quantum interference of the inner channel at $\nu_{\rm{B}}=2$ (same configuration as in Figs. \ref{Fig1}a-b). The Fourier transform of these oscillations in a small sliding window gives the plunger-gate dependence of the oscillation frequency\cite{deprez2021}.
The resulting gate-spectroscopy shown in Fig. \ref{Fig1}h reveals three peaks of decreasing amplitudes (see inset) that relate to the first harmonic frequency $f_{\rm{pg}}$ and the next two harmonics $2f_{\rm{pg}}$ and $3f_{\rm{pg}}$. This indicates quantum interference occurring over two and three turns of the inner channel loop, thus providing a clear signature of the interferometer's high coherence.
Importantly, each peak diverges at the same plunger gate voltage $V_{\rm{pg}}^{\rm c}=-0.28$ V, which corresponds to the expulsion of the inner channel from under the gate when the filling factor under the gate reaches $\nu_{\rm{pg}}\sim 1$. This divergence is channel-specific\cite{deprez2021} and provides an unambiguous indicator of the QH edge channels involved in the interference.
	
One new aspect of our measurement approach is the systematic acquisition of IV curves at every point of the interference pattern in Figs. \ref{Fig1}c,g, enabling us to simultaneously explore the complete parameter space of energy, plunger gate voltage, and magnetic field (see the plunger gate-dependent oscillations at various bias voltages in Supplementary Video 1). Figs. \ref{Fig1}c,g are extracted from the bias-dependence data at zero bias. The bias voltage dependence of the oscillation frequency yields a checkerboard pattern illustrated in Fig. \ref{Fig1}e in a restricted gate voltage range, reflecting the additional phase shift acquired by the injected electrons at finite energy. This checkerboard pattern can be accurately simulated in Fig. \ref{Fig1}f, as outlined in the Methods section. In turn, this enables us to compute the Fourier transform of the oscillations at each bias voltage (see oscillation frequency dispersion at various bias voltages in Supplementary Video 2) and to extract the bias voltage dependence of each harmonics displayed in Fig. \ref{Fig1}i. The resulting oscillatory lobe structure of each harmonic is best fitted with a Gaussian decay for the energy relaxation (see Ref.\cite{deprez2021}), and provides the Thouless energy of the interferometer edge $E_{\text{Th}} = hv/L = 135\;\upmu$V (bias-oscillation period), where $L = 3.4\;\upmu\rm{m}$ is the length of the interfering channel between two QPCs and $v$ the edge-excitation velocity. The harmonics then yield $E_{\text{Th}}/N$ (see Fig. \ref{Fig1}i inset), where $N$ is the respective harmonic index, providing an assessment of $v = 1.1\times 10^5$ $\rm{m}.\rm{s}^{-1}$ consistent with our previous work\cite{deprez2021}.

\bigskip
\textbf{Electron pairing on two coherently coupled channels}
\bigskip

\begin{figure*}[!ht]
\includegraphics[width=14cm]{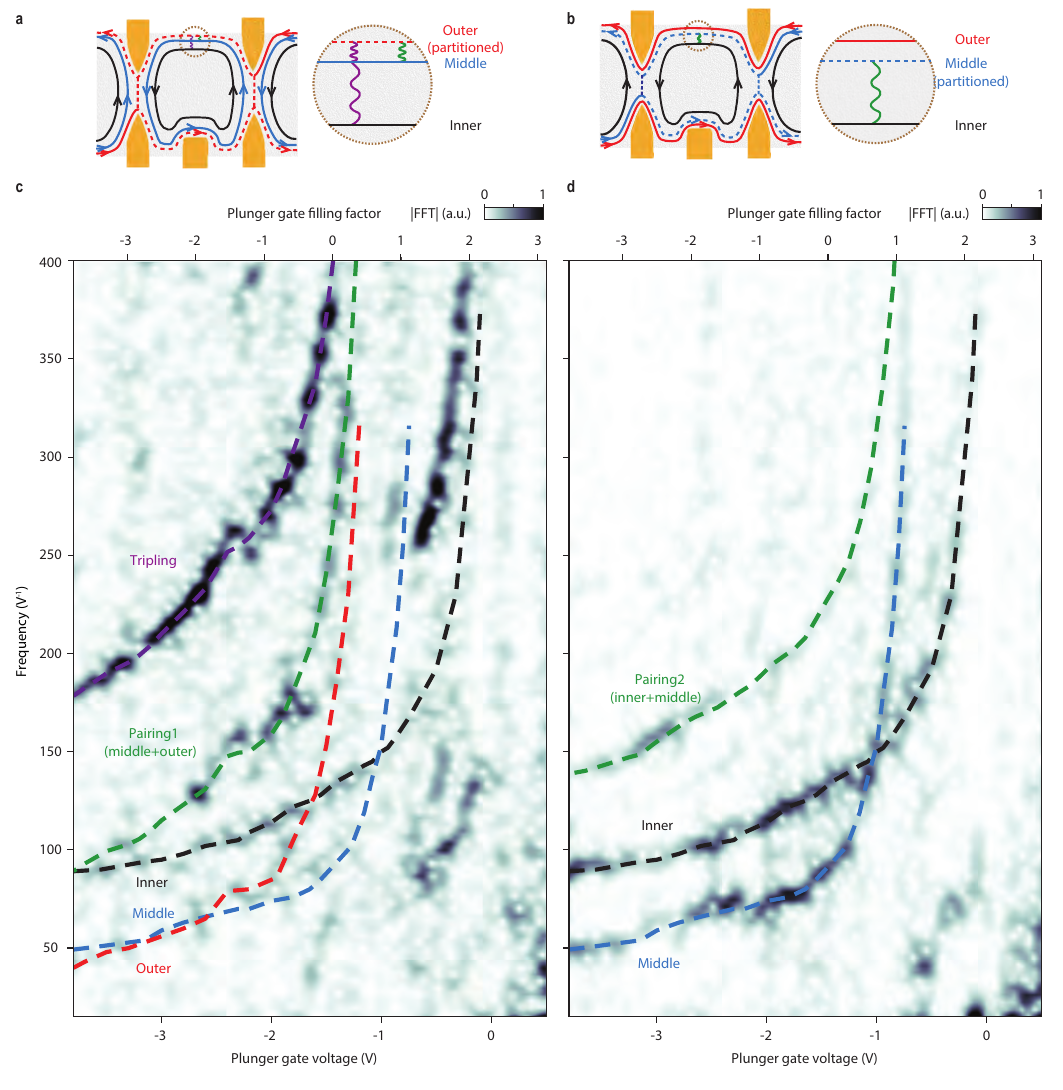}
\centering
\caption{\textbf{Tripling and pairing interference.} 
\textbf{a} and \textbf{b}, Schematic illustrations of the edge channel configurations when interfering with the outer channel in \textbf{a} and middle channel in \textbf{b}. The inner, middle, and outer edges are represented in black, blue, and red, respectively, with dashed lines indicating the partitioned edge. The green wavy line depicts the interaction between the middle and outer edges, while the violet wavy line illustrates the interactions among all three edges. \textbf{c-d,} 
Gate-spectroscopy derived from the conductance oscillations of the outer and middle edge channels, respectively, measured at $V_{\rm{bg}}=2.5~\rm{V}$ ($\nu_{\rm{B}} = 2.93$), the Fourier transforms are calculated over $0.45~\rm{V}$ and $0.24~\rm{V}$ windows in \textbf{c} and \textbf{d}. The top axis indicates the filling factor beneath the plunger gate. The black, blue, and red dashed lines represent the oscillation frequencies of the inner, middle, and outer edges. In panel \textbf{c}, the measurement is conducted with QPC transmission of $T_{\rm{1}} = 0.59$ and $T_{\rm{2}} = 0.5$. The green dashed line (pairing) is the frequency sum of the middle and outer edges, while the violet dashed line (tripling) is the sum of all three channel frequencies. In panel \textbf{d}, the measurement is performed with QPC transmission of $T_{\rm{1}} = 0.78$ and $T_{\rm{2}} = 0.81$. The green dashed line corresponds to the frequency sum of the inner and middle edge channels. Panels \textbf{c} and \textbf{d} are extracted at bias voltages $V_{\rm{DC}}=0~\rm{V}$ and $V_{\rm{DC}}=-55~\upmu \rm{V}$, respectively, from the full set of their bias voltage dependence provided in Supplementary Videos 5 and 6. Their lobe structures can be found in Supplementary Information Fig.S3. We associate the strong signal emerging between plunger gate filling factors 1 and 2, above 250 $\rm{V}^{-1}$, in panel \textbf{c} with electron tripling. The weaker signals at low frequencies ($\sim 100\,\rm{V}^{-1}$) relate to the outer and pairing (sum of the middle and outer edge channels). In this regime, the inner edge has been already expelled from beneath the plunger gate, while the middle and outer channels propagate beneath the plunger gate and are slightly pushed away from the graphene crystal edge (see detailed discussion in the Supplementary Information section VI).}
\label{Fig4}
\end{figure*}

Electron pairing emerges in our interferometer in the presence of two edge channels by interfering with the outer channel while keeping the inner channel localized in the interferometer cavity (see inset schematics in Fig. \ref{Fig2}a). Gate spectroscopy shown in Fig. \ref{Fig2}a reveals the pairing frequency (green dashed line) that is almost twice that of the inner channel interference unveiled previously in Fig. \ref{Fig1}e. Since an area variation of one flux quantum at a fixed magnetic field) is $\Delta A = \phi_0/B=\alpha \Delta V_{\text{pg}}= \alpha / f_{\rm pg}$, where $\alpha $ is the (non-linear) lever arm of the gate and $\Delta V_{\text{pg}}= 1/f_{\rm pg}$ the plunger-gate oscillation period, a frequency doubling therefore signals an abnormal flux periodicity of $h/2e$ similar to that reported in GaAs\cite{choi2015,sivan2018, Biswas2023}. 
	
The frequency doubling is also evidenced by the presence of a residual peak at half the frequency highlighted with a red dashed line that coincides with the frequency of the outer channel $h/e$-periodic interference, the latter being independently characterized by its spectroscopy at a different filling factor where pairing is sub-dominant (see Extended Data Figure 2). Inspecting the plunger-gate evolution of those frequencies, we see that both pairing and outer channel frequencies diverge at $V_{\rm pg}^{\rm c} = -0.96$ V, a value corresponding to a filling factor underneath the plunger gate $\nu_{\rm pg} = 0.05$, in agreement with the expulsion of the outer channel from the plunger gate area.

\begin{figure*}[!ht]
\includegraphics[width=12cm]{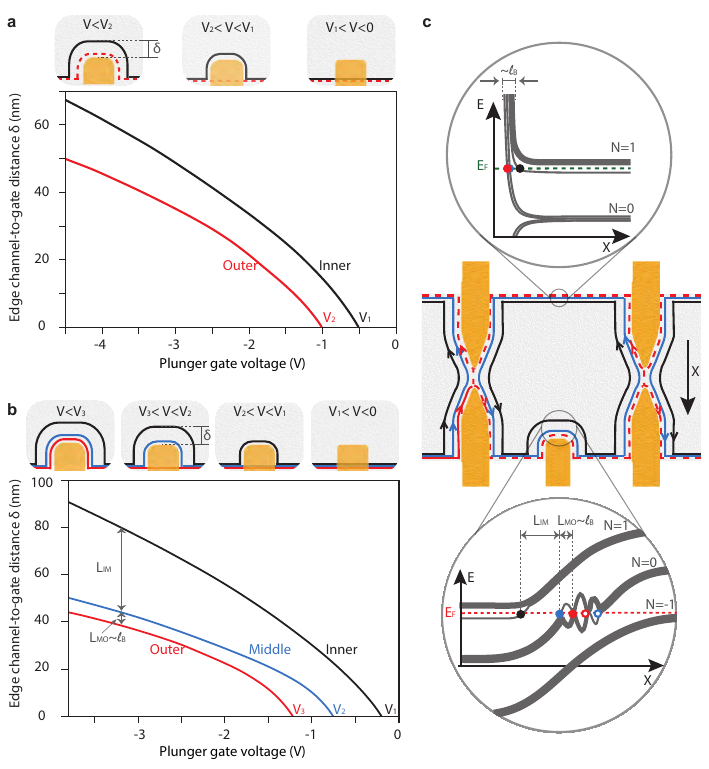}
\centering
\caption{\textbf{Interferometric determination of edge channel-to-gate distance.} 
{\textbf{a,} The distance $\delta$ between edge channel-to-plunger gate as a function of plunger gate voltage for two edge channels at $V_{\rm bg} = 1.8 \rm{V}$, extracted from the plunger gate spectroscopy shown in Fig. \ref{Fig3}a. The top schematics illustrate the edge channel configurations in three different plunger gate regimes. Calculation details are provided in the Methods section.
\textbf{b,} Edge channel-to-plunger gate distance $\delta$ for three edge channels at $V_{\rm bg} = 2.33 \rm{V}$, obtained by integrating each plunger gate frequency dispersion in Fig. \ref{Fig4}c. A distance spatially separates the outer and middle channels noted $L_{\rm{MO}}$ that is less than a magnetic length $l_B$, whereas the inner is at a distance $L_{\rm{IM}}\sim 35$ nm from the middle due to the large cyclotron gap between the zeroth and first Landau levels.} \textbf{c,} The center schematic illustrates edge configuration with a partitioned outer edge channel, represented by the dashed red line, while the black and blue lines denote the inner and middle edges, respectively. The FPI is defined by gates surrounding the quantum point contact and plunger gate (color-coded in yellow), with the crystal edge delineating the remainder. The top and bottom schematics illustrate the energy dispersion of the Landau levels along the crystal-defined and gate-defined edges, highlighting noticeable differences in edge positions. Previous studies (Refs.\cite{coissard2023, Vignaud2023}) have assessed the crystal edge dispersion, revealing edge channels confined to a few magnetic lengths from the crystal edge. At the pn junction, gaps of broken symmetry states open in the zeroth Landau level when the local filling factor $\nu_{\rm{B}}(x)$ reaches every quarter filling, i.e., $\nu_{\rm{B}}(x) = -1$, $0$, and $1$ (Refs.\cite{liu2022, Coissard22}).}
\label{Fig5}
\end{figure*}

However, unlike the harmonics in Fig. \ref{Fig1}h, the pairing frequency is not exactly twice that of the outer channel: At $V_{\rm pg} = -4.8$ V, one finds $105$ $\rm{V}^{-1}$ and $45$ $\rm{V}^{-1}$, respectively. To understand this discrepancy we add on the gate-spectroscopy the inner channel frequency (first harmonic measured in Fig. \ref{Fig1}h) as a blue dashed line in Fig. \ref{Fig2}a, which leads us to a central finding of this study: The pairing frequency is not double the AB frequency but the sum of the distinct inner and outer channel frequencies. This is seen with the green dashed line in Fig. \ref{Fig2}a, which is constructed from the sum of the black (inner channel) and red (outer channel) dashed lines, and which fits remarkably well with the pairing frequency dispersion. Here the different frequencies for the inner and outer channels stem from their slightly different positions with respect to the plunger and QPC split gates, quantified in Fig. \ref{Fig5}a (see Methods), and thus their different effective areas.

This finding is particularly striking and insightful as it demonstrates that, although localized, the inner channel influences the quantum interference of the outer channel and this unusual pairing frequency. The system therefore behaves as if a correlated excitation propagates on both inner and outer channels, thereby accumulating the sum of the AB phases of both channels $\varphi = \frac{h}{e}\oint_{inner} \textbf{A.\text{d}l} + \frac{h}{e}\oint_{outer} \textbf{A.\text{d}l}$, where $\textbf{A}$ is the vector potential.
	
Examining the bias voltage-dependence of this gate-spectroscopy displayed in Fig. \ref{Fig2}b shows that the pairing mode and the outer channels exhibit nearly the same bias voltage periodicity, that is, Thouless energy, confirming the fact that the pairing frequency is not a harmonic of the outer channel interference. Here, it is evident that pairing prevails over the $h/e$-periodic contribution from the outer channel. 

We delve further into the evolution of pairing with respect to changes in filling factor and bias voltage in Extended Data Fig.2. At a low filling factor ($V_{\rm bg} = 1.2~\rm{V}$, $\nu_{\rm{B}} = 1.7$), pairing is present but with notably lower prominence at all bias voltages (Extended Data Figs.2a,d). However, pairing significance increases and becomes dominant as the filling factor rises (Extended Data Figs.2b, c, e,f). The influence of bias voltage on the relative weight of each frequency becomes particularly noticeable when pairing and outer frequencies have similar amplitudes, as observed at $V_{\rm bg} = 1.58~\rm{V}$ ($\nu_{\rm{B}} = 2$, Extended Data Fig.2e).

Notably, the zero-bias pajama map depicted in Fig. \ref{Fig2}c displays a distinct discontinuity in the tilted AB constant phase lines, deviating from the standard pattern shown in Fig. \ref{Fig1}b. The Fourier transform reveals two contributions associated with the pairing frequency and the outer edge frequency, as shown in the inset of Fig. \ref{Fig2}c.

 Taking into account these two oscillation frequencies, a simple frequency beating can describe the observed regular discontinuity. Fig. \ref{Fig3} demonstrates how to reproduce such discontinuities in pajama maps obtained from the outer edge interference at $V_{\rm{bg}} = 1.2~\rm{V}$ ($\nu_{\rm{B}} = 1.7$), as illustrated in Fig. \ref{Fig3}c. 
 
We first simulate in Fig. \ref{Fig3}b the oscillation of the inner channel data shown in Fig. \ref{Fig3}a. In Fig. \ref{Fig3}c, a distinct negative constant phase line (see black arrows) correlates with the frequency of the bare outer channel, allowing for the extraction of its frequency, which we simulate in Fig. \ref{Fig3}d. We then compute a signal whose frequency is the sum of the inner and outer ones in order to simulate the pairing contribution in Fig. \ref{Fig3}e. 
 Adding now the pairing contribution and the outer contribution as occurring in the interferometer leads to Fig. \ref{Fig3}f, in which a beating pattern naturally emerges and remarkably fits the data Fig. \ref{Fig3}c.

Details regarding methods and simulations for other filling factors can be found in the Supplementary Information Figs. S5-6. Additionally, differences in the relative weight of pairing at various bias voltages result in diverse beating patterns and apparent phase jumps \cite{werkmeister2023}, as shown in Ext. Data. Figs 2 j-l. This bias-dependent frequency superposition is also visualized in Supplementary Video 4.

Similarly, the apparent complexity of the checkerboard patterns can also be fully replicated by summing those of the pairing mode and the bare outer edge, using the relative amplitude of each component extracted at zero bias. Four checkerboard patterns, shown in Extended Data Figs. 3a, c, e, and g, are obtained by partitioning the outer channel at different filling factors on the $\nu_{\rm{B}} = 2$ and $\nu_{\rm{B}} = 3$ plateaus (i.e. pairing and tripling regime). Employing the functional form described in the Methods but considering only the first harmonic, we have successfully simulated the checkerboard patterns seen in Extended Data Figs. 3b, d, f, h and l.

\bigskip
\textbf{Electron tripling on three coherently coupled channels}
\bigskip

The observation of pairing naturally raises the question of whether the inclusion of an additional third channel could lead to a threefold increase in frequency, namely, electron tripling, although this has not been observed thus far in GaAs\cite{choi2015,sivan2018,nakamura2019, Biswas2023}.	

To address this question we set our QH FPI to bulk filling factor 3 and, akin to the case of pairing, we partitioned the outer channel, while having the middle and inner channels localized (Fig. \ref{Fig4}a). Fig. \ref{Fig4}c reveals a new frequency in the plunger gate spectroscopy, highlighted with the violet dashed line, which is almost three times higher than that of the outer channel: At $V_{\rm pg} = -3.9 ~\rm{V}$, one finds $179$ $\rm{V}^{-1}$ and $45$ $\rm{V}^{-1}$, respectively. By overlaying the spectral dispersion of the inner (black), middle (blue), and outer (red) channels, each separately identified, we can calculate the sum of the three. The violet dashed line denotes this sum and significantly aligns with the tripling frequency. As for the case of pairing, the tripling frequency therefore results from the sum of the three distinct Aharonov-Bohm phases of the three edge channels, each characterized by a different effective area.
	
The coherent mixing and contributions of the three channels result in a pajama pattern shown in Supplementary Video 7 that is even more complex than that for the pairing. Importantly, a non-negligible pairing contribution shown in the green dashed line in Fig. \ref{Fig4}c remains present and comes from the sum of the outer and middle channel frequencies. This suggests that pairing occurs only between the partitioned channel and the nearest neighbor channel, the middle one in this case. To ascertain this deduction, we have carried out another gate-spectroscopy in a different configuration in which we partitioned the middle channel, fully transmitted the outer channel, and localized the inner channel (see Fig. \ref{Fig4}b). The resulting spectroscopy displayed in Fig. \ref{Fig4}d clearly shows a pairing contribution coming from the sum of the middle and inner channels frequencies, confirming that this pairing occurs between partially transmitted and fully localized nearest neighbor channels. 
	
 Interestingly, the amplitude of the pairing peak is weaker in this configuration (also confirmed at slightly different filling factors in Fig. S4b), while the middle and inner frequencies are visible. Here, the difference with the previous configurations at filling factor 2 is that the pairing involves channels belonging to two different Landau levels, that is, the zeroth and first ones. Consequently, the edge channels are more spatially separated due to the large cyclotron gap compared to the case of pairing between the outer and middle edges which both belong to the zeroth Landau level. The fact that pairing intensity increases with smaller separations between channels strongly suggests that inter-channel Coulomb interactions play a crucial role in pairing and tripling.

A key question to assess the Coulomb interaction quantitatively is the real space distance between edge channels. As in optical interferometry, our gate-spectroscopy provides a very accurate measurement of the interfering path, which can in turn lead to the edge channel-to-gate distance $\delta (V_{\rm pg})$ by integrating the $V_{\rm pg}$-dependence of an edge channel frequency (see Methods). The integration of frequency dispersion in Fig. \ref{Fig3}a provides the representative edge distance between two channels at $\nu_{\rm{B}}=2$, as shown in Fig. \ref{Fig5}a. Additionally, Fig. \ref{Fig5}b displays the distances among the three edges at $\nu_{\rm{B}}=3$. Strikingly, despite a relatively smooth electrostatic potential around local top gate, the outer (red line) and middle (blue line) channels are very close, with a distance {$L_{\rm{MO}}$} comparable to or even smaller than the magnetic length $l_B=\sqrt{h/2\pi eB}$, implying strongly interacting channels. For instance, we obtain {$L_{\rm{MO}}= 5.6 $} nm at $V_{\rm pg} = -3$ V with a magnetic length $l_B=6.7$ nm at $B=14$ T. On the other hand, the inner channel (black line) that belongs to the first Landau level is located at {$L_{\rm{IM}}$} $\sim 35$ nm (at $V_{\rm pg} = -3$ V) away from the middle channel. This distance can be accounted for by the large cyclotron gap between the zeroth and first Landau levels of graphene. On the contrary, along the graphene crystal edges of the FPI, the QH edge channels are known to be all confined on a scale of the order of $l_B$ to the crystal edge\cite{coissard2023, Vignaud2023}. We can thus outline the spatial structure of edge channels as well as the many-body Landau level spectra at the crystal edges and the pn junctions in our FPI with the schematics in Fig. \ref{Fig5}c. This provides unprecedented and accurate measurements of the inter-channel distances, which are crucial for a further theoretical assessment of Coulomb interactions between the channels. It is also consistent with our observation and interpretation of weaker pairing between the middle and inner channels at $\nu_{\rm{B}} = 3$ discussed above.

\bigskip
\textbf{Discussion}
\bigskip

Inter-channel interactions typically result in fractions of the electron charge being redistributed in edge magnetoplasmons \cite{Safi1999,Pham00,Levkivskyi2008,Berg09,Inoue2014}. For two co-propagating channels with strong mutual interactions (or with equal edge velocities), electrons decompose into fast (charged) and slow (neutral) plasmonic modes, each evolving in both channels. Their dynamics, solved through the matrix scattering approach \cite{Safi1999} for edge magnetoplasmons, doesn't give rise, according to Ref.\cite{Ferraro2017} to any dominant multiple electron tunneling, whose contribution to dc transport is shown to vanish at $\nu=2$ and $\nu=3$. A more recent theoretical work has taken into account charge discreteness  and has considered the limit where all edges are totally pinched off at the QPCs \cite{Frigeri2020}. It has predicted that an electron entering the FPI generates neutral plasmons, so-called neutralons, which are absorbed by a second tunneling electron; this exchange of neutralons induces a dynamical attractive interaction, thus enhancing the average Fano factor, which could potentially reach $2$. In addition to the assumption of strong pinched QPCs, this enhancement requires non-universal interaction parameters, thus couldn't yet explain the robustness of pairing observed in various experimental contexts. In addition, correlated tunneling of three electrons or more is neglected as they are less probable, which seems plausible at higher filling factors as well. Note that at $\nu = 3$ one expects strong enough inter-edge interactions to induce two neutral modes that could enhance neutralon exchange, thus pairing dynamical attraction. The present observation of tripling at $\nu=3$ doesn't favor such a mechanism, and can therefore, more generally, provide a selective criterion for theoretical explanations.

A possible scenario would consist of two or three electrons entanglement mediated by inter-edge interactions. It was shown in Ref.\cite{Burkard2000} that two electrons entangled through a double dot tend to bunch in the singlet state, leading to a doubled Fano factor, but to antibunch in the triplet state. In our setup, distinguishable edge states and valley components could offer alternative or complementary degrees of freedom to the spin. 

Besides, it is worth mentioning a recent report on pairwise electron tunneling into large quantum dots\cite{Demir2021}, reminiscent of the FPI configuration. In this context, some theoretical models predict a possible attraction --pairing-- of localized electrons resulting from the minimization of the screened Coulomb interaction\cite{Raikh1996}, and even three electrons bunching in very specific configurations\cite{putnam2021}. 
	
A different explanation of the frequency doubling and tripling based on charging effects\cite{halperin11,sivan2016, werkmeister2023}, which has long obscured Aharonov-Bohm (AB) interferometry\cite{Zhang2009}, certainly deserves careful consideration. For strong capacitive coupling between the interfering edge and the bulk, designated by the Coulomb-dominated (CD) regime, and similarly to a single-electron transistor, the conductance should not depend on the magnetic field and rather oscillates with the gate voltage which changes the occupancy of the device, leading to vertical lines in the $B,V_{G}$ plane. Nonetheless, increasing the magnetic field decreases the loop's area so that charging energy is minimized, leading to a positive slope of the equal-phase lines. The area is restored each time the flux varies by $\phi_0$ as the number of electrons on the localized states change by an integer number, leading to phase jumps as well. While the latter could be observed in intermediate regimes between the AB and CD limits\cite{halperin11}, this is not the case in the extreme CD regime as they reduce to multiples of $2\pi$, so that the interfering phase varies continuously according to the relation $\theta=-2\pi \nu\phi/\phi_0$ (thus is doubled or tripled at $\nu_{\rm{B}}=2$ or $\nu_{\rm{B}}=3$ respectively). In our FPI at $\nu_{\rm{B}}=2$, we have observed, for the inner channel, a noteworthy crossover from the CD regime to the AB regime, with an increase in the filling factor on the $h/2e^2$ quantum Hall plateau, as illustrated in Extended Data Figure 1. Nonetheless, the pairing phenomenon gets simultaneously weaker (stronger) when the inner channel is CD-dominant (AB-dominant) (see Extended Data Figure 1 a, b, c, d). Moreover, a larger QH FPI with a size of $15\,\upmu m^2$ exhibits pairing exclusively at $\nu_{\rm{B}}=2$, without the presence of the inner or outer channel frequencies (see Fig. S2). In that case, no signature of the CD regime is observed, most likely due to a smaller charging energy. In addition, the theoretical analysis in Ref.\cite{halperin11} is valid in presence of edge reconstruction, so that highly dense localized states in compressible stripes are close enough to the edges to feel strong edge-bulk interactions. Nonetheless, we have shown the absence of edge reconstruction at the crystal edges of our devices, thus pointing towards an incompressible bulk  \cite{coissard2023}. Although the question remains open, these concordant observations suggest that the CD regime is not related to the pairing and tripling phenomena, in agreement with the conclusions of Ref.\cite{choi2015}. Beyond the edge-bulk interactions, interactions between the chiral edges need to be more carefully considered and could lead us to revisit the interpretation of such a transition \cite{frigeri2019}. 
	
A concurrent work with a similar observation of frequency doubling proposes an explanation based on phase shifts induced by the discrete addition of charges in the inner channel strongly coupled to the outer one \cite{werkmeister2023}. 
The key idea is that the charge on the outer edge, which controls the FPI phase $\theta$, is itself dependent, through inter-edge coupling, on the inner edge charge. In this model, the ground-state energy variations are expanded with respect to those of the charges on both channels. Thus, contrary to \cite{Ferraro2017,frigeri2019}, the plasmonic modes are ignored; this could be justified for a short enough inner edge where electron-hole pairs creation costs high energy. This also differs from modeling interactions by Gaussian fluctuating phase in order to fit the observed voltage dependence of the AB oscillations in Fig \ref{Fig1} a.
In principle, one could generalize the same argument to $\nu_{\rm{B}}=3$. Nonetheless, we don't attribute the observed discontinuities in the pajama to phase jumps renormalized by inter-edge interactions, but rather to beating between AB oscillations of the outer channel and a pairing mode. The way these discontinuities vary with the bias voltage (see Supplementary Video 4) would also question their emergence from a charge addition mechanism.

In addition, it is not yet established that a pure electrostatic approach could account for the electron transfer 2 by 2 observed in large quantum dots\cite{Demir2021} and for shot noise measurements in interferometers based on GaAs devices\cite{choi2015, Biswas2023}. Indeed for symmetric ac voltages at two QPCs, the Fano factor is given universally by the charge of the dominant tunneling process, independently on the form, strength and range of interactions, but can be increased by non-equilibrium excited states \cite{safi_2020} (as in Ref.\cite{frigeri2019}). The argument in Ref.\cite{werkmeister2023} is also based on a well-defined charge number for the outer edge, which is questionable for almost open QPCs for which pairing has been observed as well \cite{Biswas2023}. Therefore, correlated phenomena beyond this electrostatic approximation might enter into play in this electron pairing and tripling observed in QH interferometers. 

\section*{Methods}

\subsection*{Sample fabrication}

The hBN-encapsulated graphene heterostructure was assembled from exfoliated flakes using the van der Waals pick-up technique\cite{Wang2013} and deposited onto a graphite flake serving as the back-gate electrode. The substrates are highly doped Si wafers with a $285$~nm thick SiO$_2$ layer. The flake thicknesses are 27 nm for the graphite, 45.5 nm for the bottom hBN, and 27.5 nm for the top hBN. Contacts and electrostatic gates were patterned using e-beam lithography, and Cr/Au was deposited for the contacts after etching the heterostructure with a CHF$_3$/O$_2$ plasma. Pd was deposited for the electrostatic gates, preceded by a slight O$_2$ plasma etching to remove resist residues on hBN and ensure a homogeneous electrostatic potential beneath the gate.

\subsection*{Measurements}

All measurements were performed in a dilution fridge with a base temperature of $0.01$~K at 14T. The measurement setup and filtering are described in\cite{deprez2021} and\cite{Vignaud2023}.
Systematic current-voltage characteristics were measured in a four-terminal configuration as illustrated in Fig. 1a with an acquisition card (NI-6346 from National Instruments). IV curve takes about 10 s, with oversampling enabling us to average about 1000 samples per data point. The diagonal voltage drop across the interferometer was measured with a differential FET amplifier (DLPVA-100-F-D from Femto GmbH).
A homemade multichannel 20-bit digital-analog converter (DAC) was used to adjust the various gate voltages, with noise levels below 7.5 nV/$\sqrt{\rm{Hz}}$, and a long time resolution of 1 ppm. The DAC electronic includes an ultra-stable voltage reference LTZ1000 from Linear Technology. 
Differential resistance data were obtained by numerically differentiating the current-voltage characteristics. 

\subsection*{Checkerboard pattern with harmonics}

In the presence of a single harmonics numbered $n$, the oscillation dependence with bias voltage can be described by the functional form \cite{deprez2021}:
\begin{align}
  \label{eq-single-freq}
  G^{\rm{osc}}_n =  A_n \Bigg[\beta \cos\left(n\times  \left( 2 \pi \frac{\varphi}{\phi_0} - \frac{2L}{\hbar v} eV\beta \right) \right)\\  + \overline{\beta}\cos\left(n\times \left( 2 \pi \frac{\varphi}{\phi_0} + \frac{2L}{\hbar v} eV\beta \right) \right) \Bigg] \exp\left( -\frac{(eV)^2}{\sigma_n^2}\right) \nonumber,
\end{align}
 where $\beta$ and $\overline{\beta}$ are asymmetry parameters describing how symmetric is the voltage drop on the two sides of the interferometer. $A_n$ is the $n$th harmonic oscillation amplitude. $eV$ is the voltage applied between the source and drain. $L$ is the length of the interfering channel between two QPCs, $v$ is the edge channel velocity, and $\varphi$ is the Aharonov-Bohm flux picked up by the electrons. The phenomenological Gaussian energy decay describing phase fluctuations of the interfering edge channel due to Coulomb interactions or the electric noise in the non-interfering edge channels \cite{roulleau2007} fits best our data. The checkerboard pattern in Fig. \ref{Fig1}e is very well reproduced by the sum of the three first harmonics $G^{\rm{osc}} =  \sum_n G^{\rm{osc}}_n$, with $\beta = 0.4$, $\overline{\beta} = 0.6$, $A_1 = 0.25 e^2/h$, $A_2 = 0.052 e^2/h$, $A_3 = 0.0125 e^2/h$, $\sigma_1 = 100 ~\upmu \rm{eV}$, $\sigma_2 = 80 ~\upmu \rm{eV} $, and $\sigma_3 = 65~\upmu \rm{eV} $, as shown in Fig. \ref{Fig1}f. 
In the presence of several edge channel contributions to the oscillations, the oscillation dependence of edge channel $i$ with bias voltage can be described by:
\begin{align}
  \label{eq-single-channel}
  G^{\rm{osc}}_i =  A_i \Bigg[\beta_i \cos\left( 2 \pi \frac{\varphi_i}{\phi_0} - \frac{2L}{\hbar v_i} eV\beta_i  \right)\\  + \overline{\beta_i}\cos\left( 2 \pi \frac{\varphi_i}{\phi_0} + \frac{2L}{\hbar v_i} eV\beta_i  \right) \Bigg] \exp\left( -\frac{(eV)^2}{\sigma_i^2}\right) \nonumber,
\end{align}
with the same notation as above. Checkerboard patterns in Extended Data Fig. 3 are very well reproduced by the sum of the outer and pairing conductance $G^{\rm{osc}} =  G^{\rm{osc}}_o + G^{\rm{osc}}_p$ without harmonics. Parameters for Extended Data Fig. 3 are given in Extended Data Table 1.

\begin{table}
	\centering
		\begin{tabular}{|c|c|c|c|c|c|c|}
		 \hline
			$V_{BG} (V)$ & $\beta_o$ & $\beta_p$ & $A_o$ ($e^2/h$) & $A_p$ ($e^2/h$) & $\sigma_o$ ($\mu$eV) & $\sigma_p$ ($\mu$eV)\\
			\hline
			$1.2$ & $0.35$ & $0.65$ & $0.1$ & $0.03$ & $80$  & $80$ \\
			\hline
			$1. 58$  & $0.9$ & $0.1$ & $0.1$ & $0.1$ & $80$  & $80$ \\
			\hline
			$1.8$ & $0.6$ & $0.1$ & $0.03$ & $0.2$ & $85$  & $100$ \\
			\hline
			$2.5$ & $0.6$ & $0.4$ & $0.03$ & $0.1$ & $70$  & $70$ \\
			\hline
		\end{tabular}
	\caption{Simulation parameters corresponding to Extended Data Fig 3.}
	\label{tab:SimulationParametersExtendedDataFig3}
\end{table}

\subsection*{Gate-to-edge channel distance from gate-spectroscopy}

The gate-spectroscopy is a direct measure of the capacitance coupling between the gate and the interfering edge channels\cite{deprez2021}. 
Assuming a distance $\delta $ between the gate and the interfering channel as drawn in the left inset in Fig. \ref{Fig5}a, the lever arm of the gate is given by $\alpha (V_{\rm pg}) = L_{\rm pg} \frac{\text{d} \delta (V_{\rm pg})}{\text{d} V_{\rm pg}}$, where $L_{\rm pg} = 1.5\;\upmu\rm{m}$ is the gate edge length\cite{deprez2021}. As a result, one can compute the displacement distance $\delta $ by integrating the gate-voltage dependence of the frequency of the considered channel: $\delta (V_{\rm pg}) =\frac{\phi_0}{BL_{\rm pg}}\int^{V_{\rm pg}}_{V_{\rm pg}^{\rm c}} f_{\rm pg}(V)\text{d}V$, using $\Delta A = \phi_0/B=\alpha \Delta V_{\text{pg}}= \alpha / f_{\rm pg}$. Fig. \ref{Fig5}b displays the resulting distances for the three channels at $\nu_{\rm{B}} = 3$ computed from Fig. \ref{Fig4}c.

\subsection{Coulomb dominated to Aharonov-Bohm dominated interference transition on the inner edge} 

We present in this section and Extended Data Fig. 1 the continuous evolution of inner edge channel interferences (no pairing in that case) from Coulomb dominated to Aharonov-Bohm type when increasing the filling factor on the $h/2e$ plateau\cite{Zhang2009}. The magnetic field periodicity of the inner edge channel interference continuously evolves as the filling factor is increased as shown in Extended Data Fig. 1b. On the left of the plateau, the oscillations are completely Coulomb dominated i.e. the period is equal to the Aharonov-Bohm period but with an opposite slope of constant phase line on the pajama plot\cite{halperin11} as shown in Extended Data Fig. 1c. On the middle of the plateau, the oscillations are almost magnetic field independent as shown in Extended Data Fig. 1d. On the right side of the plateau, the oscillations correspond to an Aharonov-Bohm modulation as shown in Extended Data Fig.~1e and Fig. 1c. This evolution is consistent with a reduced bulk-inner edge interaction as the filling factor is increased. Note that on the right part of the plateau, the Fermi level is pinned in the Lifshitz tail of the N=1 Landau level such that the bulk compressible island is separated from the edge channels by an incompressible strip of width similar to the distance between inner and middle channel assessed in Fig. 4d. On the contrary, on the left part of the plateau, the Fermi level in pinned at the top of the zeroth LL, which forms a compressible island of localized states that are potentially closer to the edge channels.

\subsection*{Evolution of electron pairing with filling factor}

The relative contributions of the outer edge channel and pairing to the interference signal vary with the filling factor when partitioning the outer edge channel on the $h/2e$ quantum Hall plateau. On the left side of the plateau, at $V_{\rm{bg}} = 1.2~\rm{V}$ ($\nu_{\rm{B}} = 1.7$), the interference signal is dominated by the outer edge channel contribution, as observed in Extended Data Figs. 2a and d. In the middle of the plateau, at $V_{\rm{bg}} = 1.58~\rm{V}$ ($\nu_{\rm{B}} = 2$), the contributions from the outer edge channel and pairing are comparable, as shown in Extended Data Fig. 2b and e. On the right side of the plateau, at $V_{\rm{bg}} = 1.8~\rm{V}$ ($\nu_{\rm{B}} = 2.26$), the pairing contribution becomes dominant, as illustrated in Fig. 2a and b. 

It should be noted that the contribution weights of the outer channel and pairing are influenced by the bias voltage. This relationship is evident in the lobe structure observed at $V_{\rm{bg}} = 1.58~\rm{V}$ ($\nu_{\rm{B}} = 2$), shown in Extended Data Fig. 2e and the corresponding pajama maps in Extended Data Figs. 2h, j, k, l.

\section*{Acknowledgments}
We thank A. Assouline, D. Basko, P. Degiovanni, M. Heiblum, B. Rosenow, K. Snizhko, and E. Sukhorukov for valuable discussions. We thank F. Blondelle for technical support on the experimental apparatus. Samples were prepared at the Nanofab facility of the N\'eel Institute. 
This work has received funding from the European Union's Horizon 2020 research and innovation program under the ERC grant \textit{SUPERGRAPH} No. 866365. B.S., H.S., and W.Y. acknowledge support from the QuantERA II Program which has received funding from the European Union's Horizon 2020 research and innovation program under Grant Agreement No 101017733. K.W. and T.T. acknowledge support from the JSPS KAKENHI (Grant Numbers 20H00354, 21H05233 and 23H02052) and World Premier International Research Center Initiative (WPI), MEXT, Japan.



\bibliography{Yang-Bib}


\setcounter{figure}{0}
\renewcommand{\figurename}{Extended Data Fig.}
\renewcommand{\tablename}{Extended Data Table}

\begin{figure*}[!ht]
\centering
\includegraphics[width=16cm]{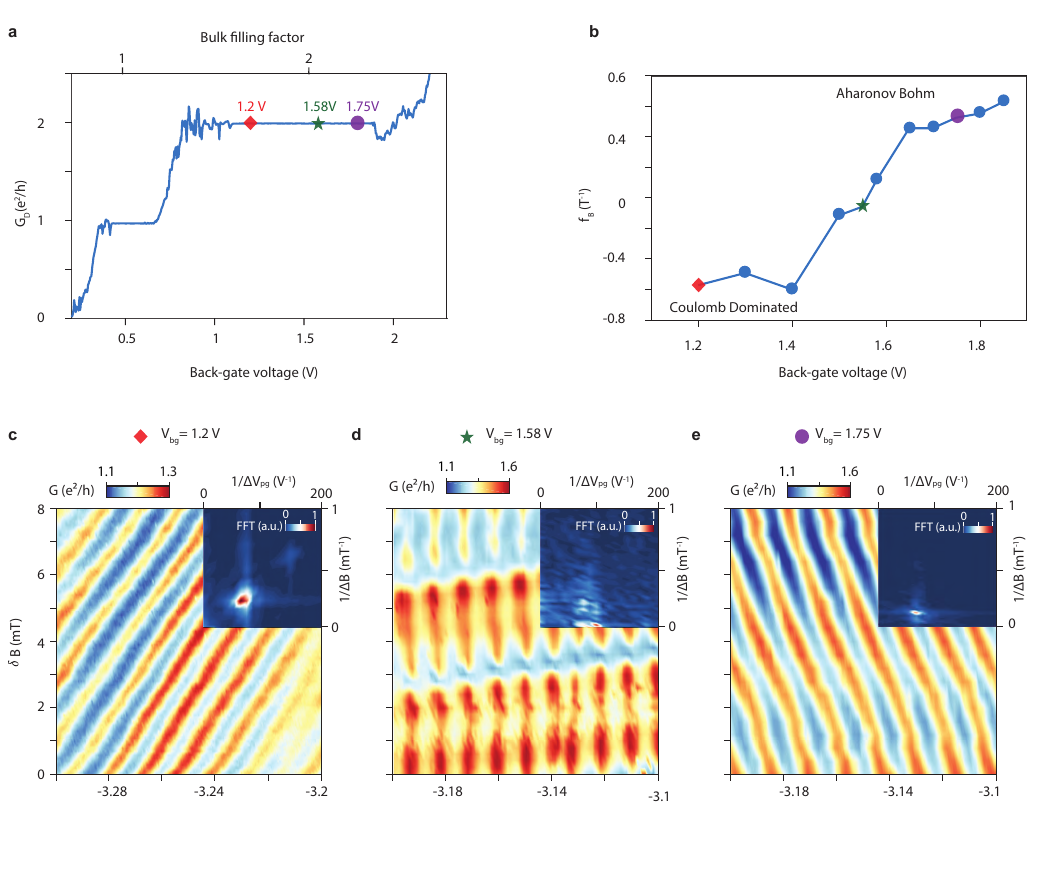}
\caption{\textbf{Transition from Coulomb dominating to Aharonov-Bohm dominating of inner edge channel interference.}
 \textbf{a}, Diagonal conductance across the interferometer as a function of back-gate voltage when both QPC are open. The color symbols represent the back-gate voltages of $1.2~\rm{V}$, $1.58~\rm{V}$, and $1.75~\rm{V}$.
 \textbf{b}, Magnetic field oscillation frequency is plotted against the back-gate voltage when pinching the inner edge channel. The positive (negative) signs indicate the constant phase lines with a positive (negative) slope in the pajama maps. \textbf{c}, Pajama obtained by pinching the inner edge channel at $V_{\rm{bg}} = 1.2~\rm{V}$, with QPC transmission of $T_{\rm{1}} = 0.68$ and $T_{\rm{2}} = 0.64$. \textbf{d.} Pajama obtained at $V_{\rm{bg}} = 1.58~\rm{V}$ by partially transmitting the inner edge channel with the transmission of $T_{\rm{1}}=T_{\rm{2}}= 0.69$. \textbf{e} Pajama obtained at $V_{\rm{bg}} = 1.75~\rm{V}$ with QPC transmission of $T_{\rm{1}} = 0.8$ and $T_{\rm{2}} = 0.6$.}
\label{ExtDataFig1}
\end{figure*}

\newpage

~\vfill
\begin{figure*}[!ht]
\centering
\includegraphics[width=12cm]{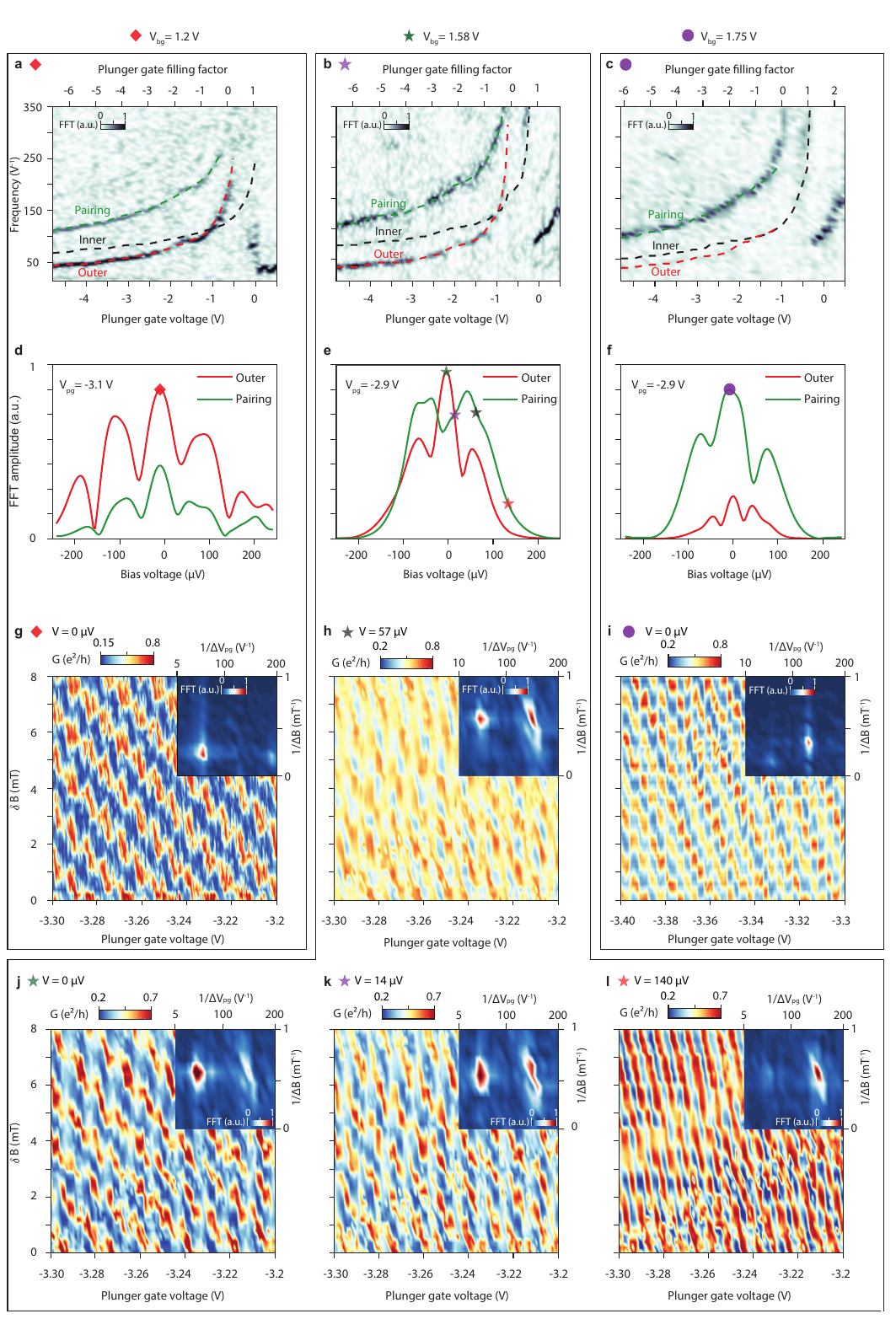}
\caption{\textbf{Electron pairing evolution with filling factor and bias voltage when partially transmitting the outer edge channel.} 
\textbf{a-c}, FT amplitude extracted from oscillations of the outer channel at $V_{\rm{bg}} = 1.2~\rm{V}$, $V_{\rm{bg}} = 1.58~\rm{V}$ and $V_{\rm{bg}} = 1.75~\rm{V}$, versus plunger gate voltage and oscillation frequency. Black dashed lines are independently derived from the inner channel interference. The dashed blue lines indicate the oscillation frequency of the outer channel. The sum of these two frequencies is computed as a dashed green line. 
\textbf{d-f}, The lobe structures of paired channels (green colored) and outer channel (blue colored) are extracted at $V_{\rm{bg}} = 1.2~\rm{V}$, $V_{\rm{bg}} = 1.58~\rm{V}$ and $V_{\rm{bg}} = 1.75~\rm{V}$, respectively. The pairing frequency dominates over the entire bias ranges at $V_{\rm{bg}} = 1.2~\rm{V}$ (\textbf{d.}), becomes comparable with the outer channel frequency at $V_{\rm{bg}} = 1.58~\rm{V}$ (\textbf{e.}), and is less evident at $V_{\rm{bg}} = 1.75~\rm{V}$ (\textbf{f.}). \textbf{g,i}, Pajama maps of the outer channel interference at $V_{\rm{bias}} = 0 ~\upmu \rm{V}$: $V_{\rm{bg}} = 1.2~\rm{V}$ with QPC transmission of $T_{\rm{1}} = 0.98$ and $T_{\rm{2}} = 0.96$  (\textbf{g.}) ; $V_{\rm{bg}} = 1.75~\rm{V}$ with QPC transmission of $T_{\rm{1}} = 0.85$ and $T_{\rm{2}} = 0.79$ (\textbf{i.} ). \textbf{h,j-l.} Pajama maps obtained at $V_{\rm{bg}} = 1.58~\rm{V}$ under various bias voltages of $V_{\rm{bias}} = 57 ~\upmu \rm{V}$ (\textbf{h.}), $0 ~\upmu \rm{V}$ (\textbf{j.}), $14~\upmu \rm{V}$ (\textbf{k.}), and $140~\upmu \rm{V}$ (\textbf{l.}). These bias voltages are marked accordingly in panel (\textbf{e.})}
\label{ExtDataFig2}
\end{figure*}

\newpage

~\vfill
\begin{figure*}[!ht]
\centering
\includegraphics[width=16cm]{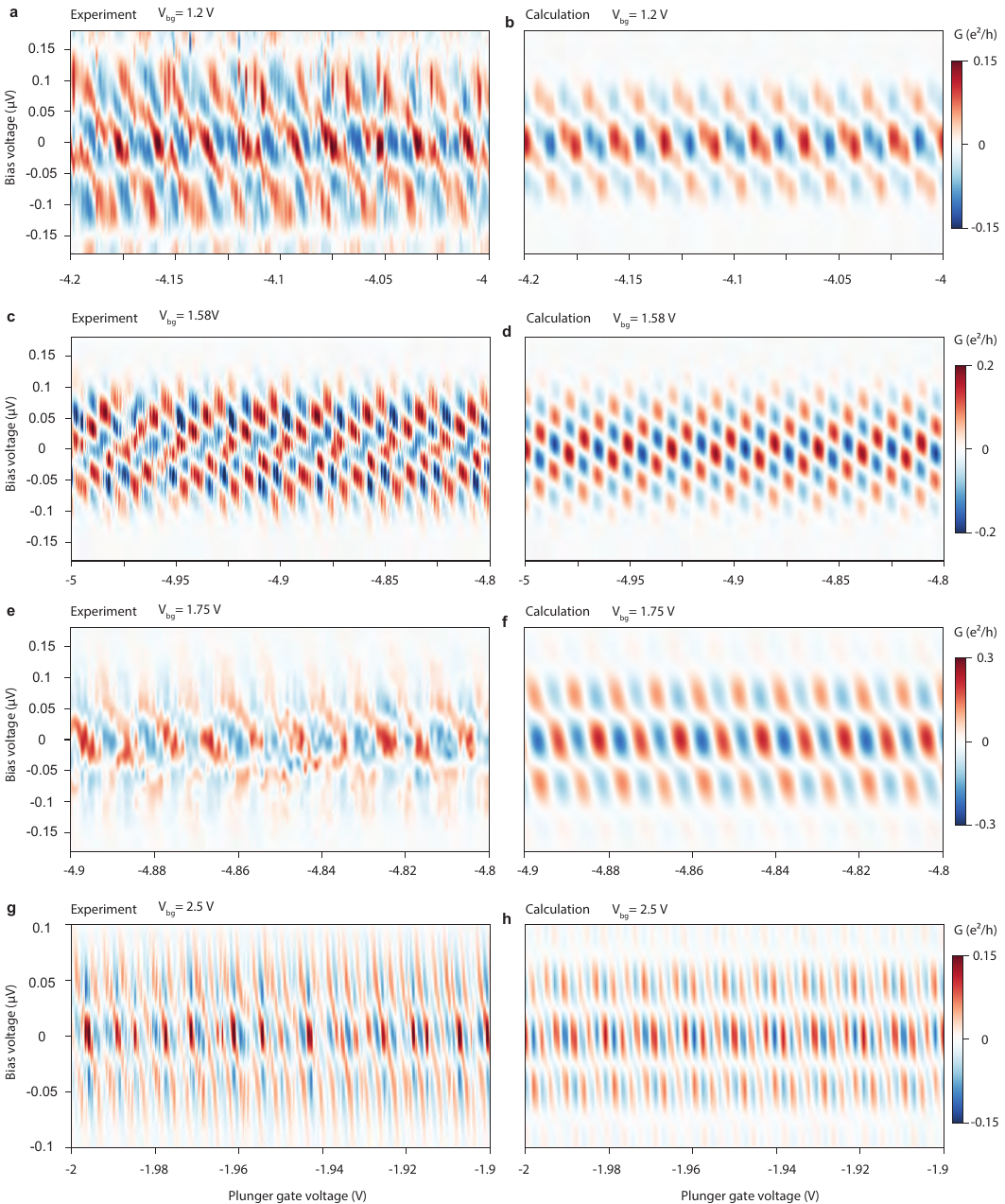}
\caption{\textbf{Experiments and simulations on checkerboard patterns of the outer edge channels under various back-gate voltages.}
  Checkerboard patterns of the outer edge channel were measured at back-gate voltages of $1.2~\rm{V}$ (\textbf{a}), $1.58~\rm{V}$ (\textbf{c}), $1.8~\rm{V}$ (\textbf{e}), and  $2.5~\rm{V}$ (\textbf{g}).
 \textbf{b-h}, Simulations were conducted to calculate the checkerboard patterns by incorporating the contribution of pairing electrons and outer edge channel for each back-gate voltage. The simulated patterns match the observed checkerboard of the outer edge channel. }
\label{ExtDataFig3}
\end{figure*}

\clearpage


\clearpage
\onecolumngrid
\setcounter{figure}{0}
\setcounter{section}{0}
\renewcommand{\figurename}{Fig.}
\renewcommand{\thefigure}{S\arabic{figure}}

\newpage

~\vspace{5em}
\part*{ \centering Supplementary Information}
\bigskip
\vspace{2em}

\section{Quantum Hall characterization}
\label{secQH}

\begin{figure}[ht!]
	\centering
		\includegraphics[width=0.6\textwidth]{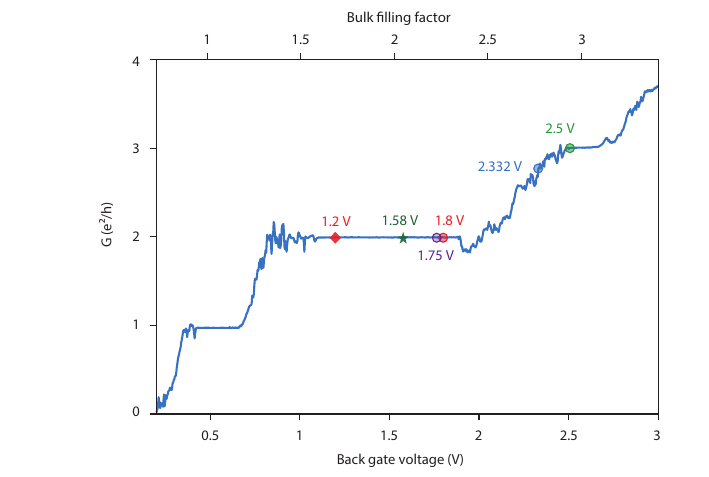}
	\caption{\textbf{Quantum Hall plateaus.} Diagonal conductance of device WY50 of the main text as a function of back-gate voltage at 14 T. Top axis is the bulk filling factor. The dots on the quantum Hall plateau indicate the back-gate voltage where each figure was measured: $V_{\rm{bg}} = 1.2~\rm{V}$ ($\nu = 1.7$, red diamond ), $V_{\rm{bg}} = 1.58~\rm{V}$ ($\nu = 2$, green star),$V_{\rm{bg}} = 1.8~\rm{V}$ ($\nu = 2.26$, red dot). For higher filling factor, we present data obtained at $V_{\rm{bg}} = 2.33~\rm{V}$ ($\nu = 2.8$, blue dot) and $V_{\rm{bg}} = 2.5~\rm{V}$ ($\nu = 2.93$, green dot).}
	\label{figHall}
\end{figure}

Here we highlight the back-gate voltages corresponding to the bulk filling factors at which the interferometry experiments were performed. Figure \ref{figHall} shows the diagonal conductance of device WY50 of the main text, measured in the configuration shown in Fig. 1a with both QPCs fully open. Integer quantum Hall states, including broken symmetry states, are well developed as featured by quantized diagonal conductance when sweeping the back-gate voltage at 14 T. We summarize below the back-gate voltages $V_{\rm{bg}}$ (indicated by colored dots in Fig. \ref{figHall}) and corresponding bulk filling factors $\nu$ for all figures of the work.

\begin{itemize}
	\item $V_{\rm{bg}} = 1.2~\rm{V}$ ($\nu = 1.7$) : Ext. Data. Figs. 1a,c and 2a,c (red diamond in Fig. S1).
	\item $V_{\rm{bg}} = 1.58~\rm{V}$ ($\nu = 2$): Ext. Data. Figs. 1b,d and 2b,d-g (green star in Fig. S1).
	\item $V_{\rm{bg}} = 1.75~\rm{V}$ ($\nu = 2.21$): Ext. Data. Figs. 1e and 2c,f-i (purple dot in Fig. S1).
	\item $V_{\rm{bg}} = 1.8~\rm{V}$ ($\nu = 2.26$) : Figs. 1-3 and Fig. S3 and Supplementary Videos 1-4 (red dot in Fig. S1).
	\item $V_{\rm{bg}} = 2.33~\rm{V}$ ($\nu = 2.8$) : Fig. S5 and Supplementary Videos 8-10 (blue dot in Fig. S1).
	\item $V_{\rm{bg}} = 2.5~\rm{V}$ ($\nu = 2.9$) : Figs. 4 and 5 and Supplementary Videos 5 and 6 (green dot in Fig. S1).
\end{itemize}

\section{Evolution of pairing strength with interferometer size}
\label{secInterferometerSize}

\begin{figure}[ht!]
	\centering
		\includegraphics[width=1.0\textwidth]{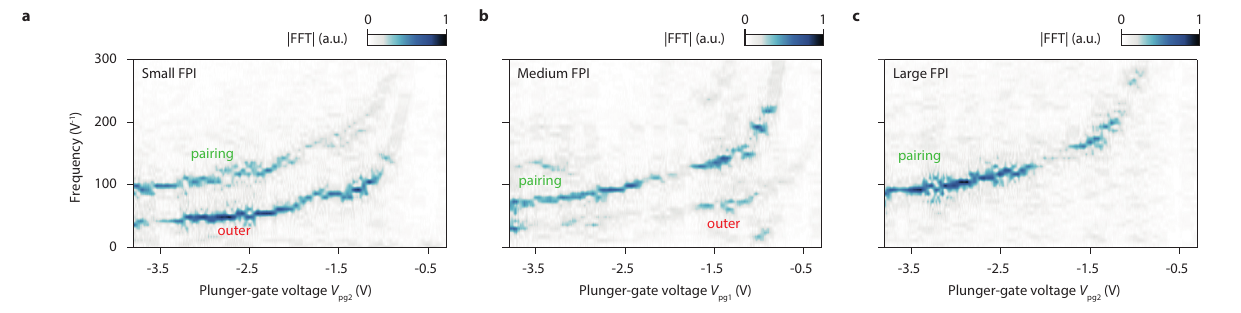}
	\caption{\textbf{Evolution of pairing strength with interferometer size}. Fourier transform amplitude of conductance oscillations versus plunger gate voltage and plunger gate frequency when partitioning the outer edge channel at $\nu=2$. \textbf{a}, Small FPI of $3.1~\upmu \rm{m}^2$ is measured at $\nu = 2.3$. \textbf{b}, Medium FPI of $10.7~\upmu \rm{m}^2$ is measured at $\nu = 2.3$. \textbf{c}, Large FPI of $14.7~\upmu \rm{m}^2$ is measured at $\nu = 2.5$.}
	\label{FigS2}
\end{figure}

Here, we present additional data obtained from sample BNGr74 comprising two interferometers in series, which were extensively studied in ~\cite{deprez2021}. This device configuration allows to study, in the same sample, three interferometers of different sizes: $3.1~\upmu \rm{m}^2$, $10.7~\upmu \rm{m}^2$ and $14.7~\upmu \rm{m}^2$. The measurements were performed with standard lock-in at zero-bias voltage. Figs. \ref{FigS2}a-c exhibit the gate spectroscopy of the outer edge channel as a function of plunger gate voltage in small, medium and large interferometers, respectively. The pairing signal is more pronounced in the medium interferometer and is present alone in the large interferometer where charging effects are expected to be lessened, indicating an anti-correlation between charging energy and the emergence of the pairing frequency.

\section{Lobe structure at $\nu=3$}
\label{secLobeStructure}

\begin{figure}[ht!]
	\centering
		\includegraphics[width=1\textwidth]{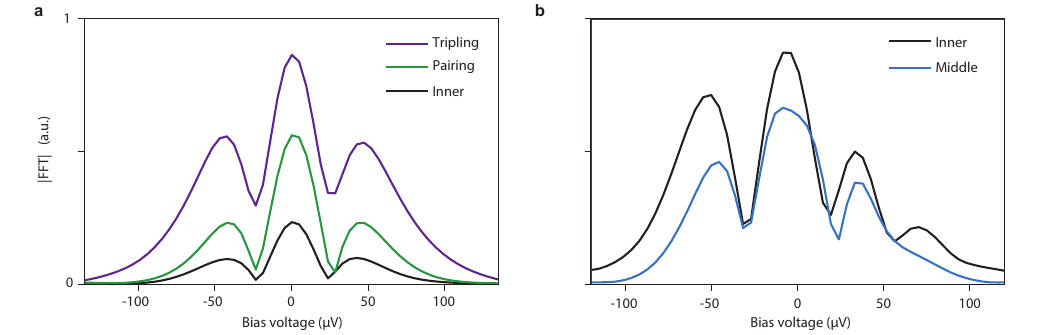}
\caption{\textbf{Lobe structure of edge channels at $V_{\rm{bg}} = 2.5~\rm{V}$ at $\nu=3$}. \textbf{a}, The lobe structures obtained when partitioning the outer edge chanel at $V_{\rm{bg}} = 2.5~\rm{V}$ shown in the main text Fig. 4c. The solid black, green, and purple lines represent the inner channel, pairing (between the middle and outer channels), and tripling, respectively. \textbf{b}, Lobe structures obtained by partitioning the middle edge channel, corresponding to the data presented in the main text Fig. 4d. The solid black and blue lines correspond the inner and middle edge channels, respectively.}
	\label{figLobeTripl}
\end{figure}

We show in Fig. \ref{figLobeTripl} the Fourier transform amplitude --lobe structure-- as a function of DC bias voltage extracted from Fig. 4. It illustrates the lobe structures associated with the pairing mode, tripling mode and inner edge channel (extracted from Fig. 4c) in  Fig. \ref{figLobeTripl}a, and those related to the inner and  middle channels (depicted in Fig. 4d) in Fig. \ref{figLobeTripl}b. Unlike the case of harmonics observed in Fig. 1i, all Fourier transform amplitudes here exhibit comparable periodicities in bias voltage. The extracted Thouless energies, of the order of $100~\upmu\rm{V}$, are comparable to that of the inner edge channel shown in Fig. 1i. We also note that the asymmetric lobes structure occurs in observed lobe structures, as illustrated in Fig3 and Extended Data Figures 2 d-f as well. In these plots, a positive (negative) bias voltage brings the edge channels closer (further), resulting in a relatively small effect that is nonetheless visible in the asymmetric lobes structure. 

\section{Tripling at another bulk filling factor}
\label{secTripling}

\begin{figure}[ht!]
	\centering
		\includegraphics[width=0.9\textwidth]{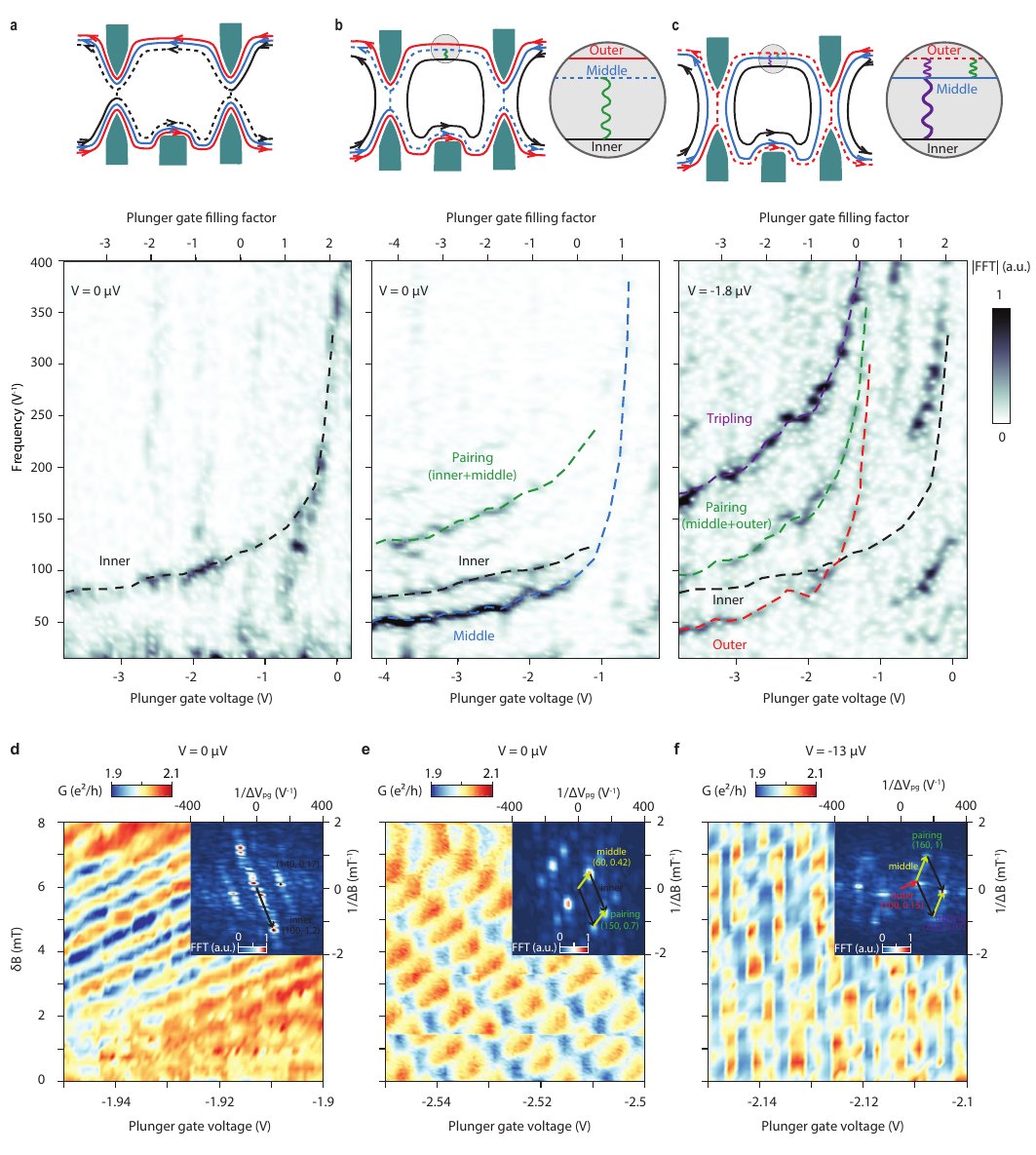}
	\caption{\textbf{Interference behavior at $V_{\rm{bg}} = 2.33~\rm{V}$ ($\nu_{\rm{B}} = 2.8$)}. \textbf{a}, The gate spectroscopy of the inner edge channel. The oscillations exhibit a single frequency. \textbf{b}, When partitioning the middle edge channel, observed frequency dispersions can be associated with the inner (dashed black line), middle channel (dashed blue line), and a weakly visible frequency sum of inner and middle (pairing mode, dashed green line). \textbf{c.} When partitioning the outer edge channel, the lowest frequency dispersion is related to the bare outer edge (dashed red line). The strongest frequency dispersion is associated with the total frequency of all three channels (tripling mode, dashed purple line). The frequency sum of the outer and middle (pairing mode) is also visible, indicated by a dashed green line. \textbf{d-f}, Pajama maps are measured by partitioning the inner, middle, and outer edges, respectively. FFT analysis is performed and presented in the inset of each pajama. The data shown are taken at zero bias from the IV curves. Supplementary videos 8-10 display the pajama maps across the entire range of bias voltages.} 
	\label{figTripleOtherNu}
\end{figure}

Fig. \ref{figTripleOtherNu} displays the interference patterns observed at $V_{\rm{bg}} = 2.33~\rm{V}$ ($\nu_{\rm{B}} = 2.8$), involving three edge channels, similar to the data presented in Fig. 4. The gate spectroscopy, obtained from conductance oscillations when partitioning the inner, middle, and outer edge channels at zero bias, is plotted in Figs. \ref{figTripleOtherNu}a-c respectively. Similarly to Fig. 4d, the emergence of electron pairing (dashed green line) between the inner (dashed black line) and middle edge channel (dashed blue line) is observed by partitioning the middle edge channel. When interfering with the outer edge channels, the dominant signal corresponds to the sum of all three channels (tripling mode, indicated by a dashed purple line), while the bare outer channel frequency (dashed red line) and pairing mode are weakly visible.

The pajama maps are measured accordingly, and the inner channel interference in Fig. \ref{figTripleOtherNu}d exhibits a CD regime for all bias voltages, as demonstrated in Supplementary Video 8. Intricate patterns involving multiple frequencies are observed when partitioning both the middle (Fig. \ref{figTripleOtherNu}e) and outer (Fig. \ref{figTripleOtherNu}f) edge channels. The precise regime for the middle channel interference is detailed in Supplementary Video 9, displaying a pattern dominated by an AB regime at high bias, attributed to the middle channel through its plunger gate periodicity, and a mixed pattern around zero bias, which can be seen as a combination of the AB dominated middle channel oscillation and a weaker CD dominated pairing mode. FFT analysis is performed for each map, revealing that tripling and pairing modes arise from the frequency sum. Supplementary Video 10 shows the pajama of the outer channel interference at all bias voltages.

\section{Simulation of phase jumps in the pajama maps}
The interruptions observed in the pajama maps of the outer edge channel, as depicted in Fig. 2c and Extended Data Fig. 2, can be reproduced through frequency beating. This effect arises when multiple frequencies contribute to the conductance oscillations at the same time. Regarding the bias voltage dependence, each channel (labeled by $l$) contribution can be written as 
\begin{align}
  \label{eq-single-channel-paj1}
  {G^{\rm{osc}}}_{l} =  A_l \cos\left( 2 \pi (f_{\rm{B},l} B + f_{\rm{pg},l} V_{\rm{pg}}) \right),
\end{align}

The pairing/tripling mode can be written as
\begin{align}
  \label{eq-single-channel-paj2}
  G^{\rm{osc}}_{pair} =  A_{pair} \cos\left( 2\pi ((f_{\rm{B},in}+f_{\rm{B},out}) B + (f_{\rm{pg},in}+f_{\rm{pg},out})V_{\rm{pg}}) \right) \nonumber, \\
	  G^{\rm{osc}}_{tri} =  A_{tri} \cos\left( 2\pi ((f_{\rm{B},in}+f_{\rm{B},mid}+f_{\rm{B},out}) B + (f_{\rm{pg},in}+f_{\rm{pg},mid}+f_{\rm{pg},out})V_{\rm{pg}}) \right) \nonumber, \\ 
\end{align}

The total conductance oscillation is computed by summing the contributions from each channel, represented as $G^{\rm{osc}} =  \sum_n G^{\rm{osc}}_n$. Figs. S5-6 display that these calculated pajama maps based on $G^{\rm{osc}}$ are in excellent agreement with the experimental data. The specific contributions from each channel are detailed in Table S1.

\begin{table}
	\centering
		\begin{tabular}{|c|c|c|c|}
		 \hline
			$V_{BG} (V)$ & $A_{out}$ ($e^2/h$) & $A_{pair}$ ($e^2/h$) & $A_{tri}$ ($e^2/h$)\\
			\hline
			$1.58$ & $0.07$ & $0.07$ & $0$  \\
			\hline
			$2.5$  & $0.16$ & $0.16$ & $0.32$  \\
			\hline
		\end{tabular}
	\caption{Simulation parameters corresponding to Figs. S5-6.}
	\label{tab:SimulationParametersSIFig3}
\end{table}

\begin{figure*}[ht]
	\centering
		\includegraphics[width=14cm]{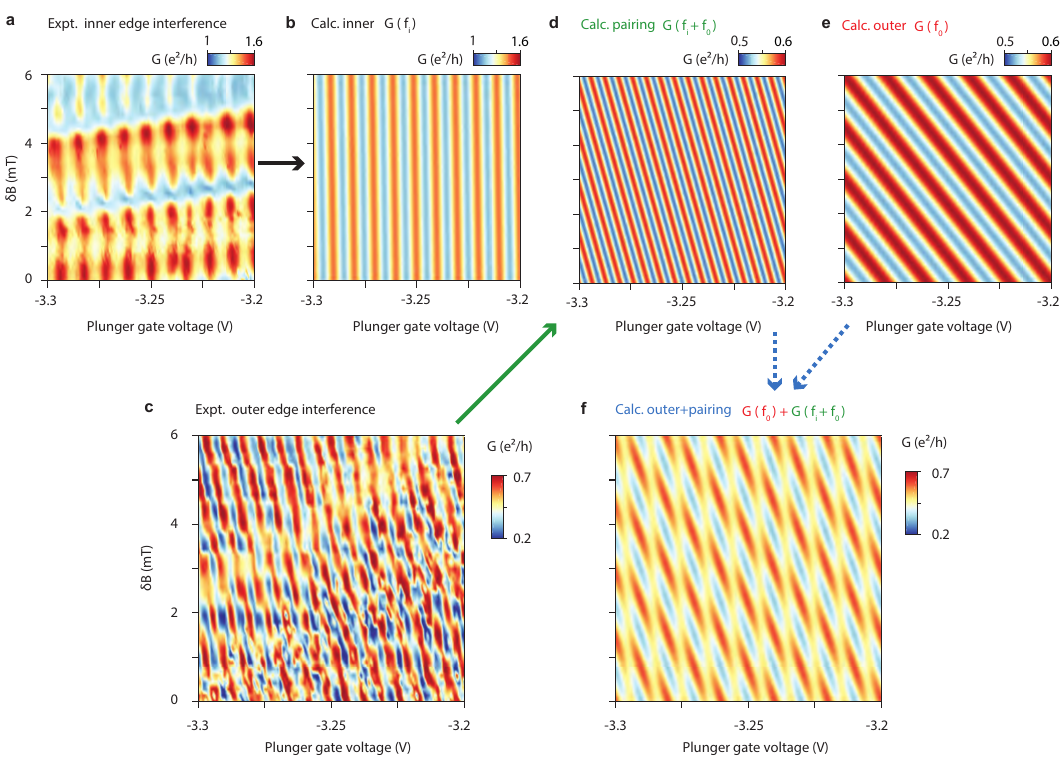}
	\caption{\textbf{Experimental and simulated pajama maps at $V_{\rm{bg}} = 1.58~\rm{V}$ ($\nu_{\rm{B}} = 2$)}. \textbf{a,} Experimental pajama map obtained by partitioning the inner edge channel, showing a single-frequency oscillation. \textbf{b,} Calculated pajama map for the inner channel interference, derived from the oscillation frequency observed in panel \textbf{a}. \textbf{c,} Experimental pajama map from partitioning the outer edge channel, where the dominated frequency is associated with electron pairing. \textbf{d,} Simulated pajama map with oscillation frequency corresponding to electron pairing, based on the dominant frequency observed in panel \textbf{c}. \textbf{e,} Pajama map for the bare outer edge channel, determined by subtracting the oscillation frequency of the inner channel from that of the pairing mode. \textbf{f,} The beating phenomenon in the pajama map, observed when oscillations from both electron pairing and the outer edge channel are present simultaneously.}
	\label{figpajSimul1p5V}
\end{figure*}

\begin{figure*}[ht]
	\centering
		\includegraphics[width=14cm]{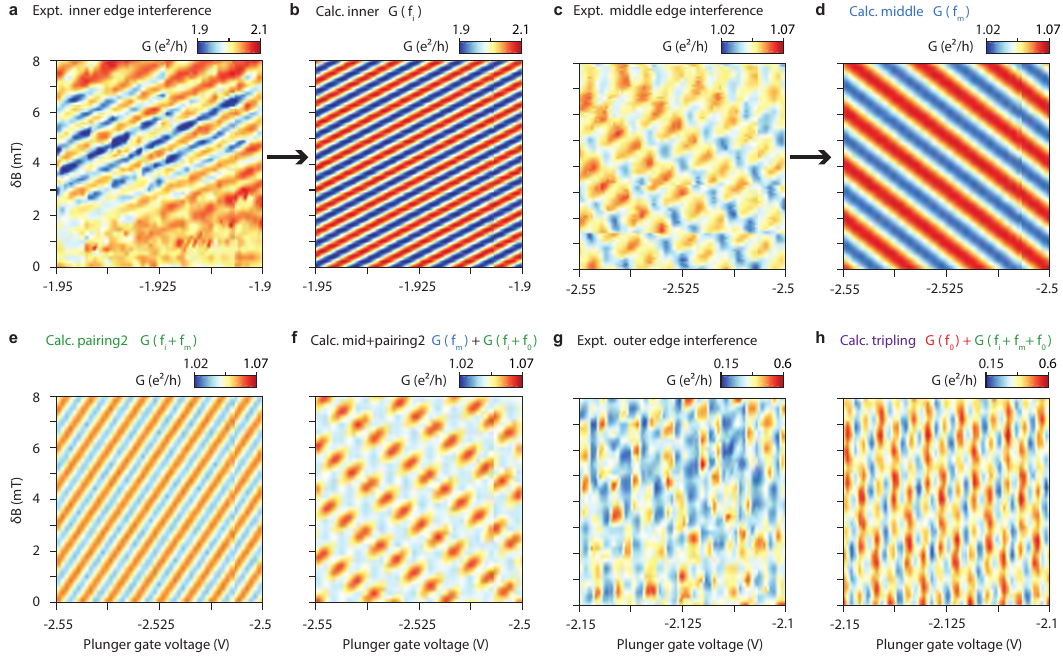}
	\caption{\textbf{Experimental and simulated pajama maps at $V_{\rm{bg}} = 2.5~\rm{V}$ ($\nu_{\rm{B}} = 2.9$)}. \textbf{a}, Pajama map obtained by partitioning the inner edge channel. \textbf{b}, The calculated pajama map for the inner edge channel, derived from the oscillation frequency observed in panel \textbf{a}. \textbf{c.} Pajama map obtained by partitioning the middle edge channel. The dominant oscillation signal originates from the middle edge channel. \textbf{d,} Simulated pajama for middle edge channel interference, based on the dominant frequency observed in panel \textbf{c}. \textbf{e,} Pajama map showing electron pairing between the inner and middle channel, deduced by summing the oscillation frequencies of the inner and middle edge channels.
    \textbf{f,} Pajama map acquired through partitioning the outer edge channel. The dominant frequency corresponds to the tripling of all channels, allowing us to extract the oscillation frequency of the outer edge channel. \textbf{g,} Simulated pajama map when oscillations from both tripling and the outer edge channel are present.}
	\label{Figure_S6}
\end{figure*}

Fig. \ref{figpajSimul1p5V} presents simulations of the pajama maps at $V_{\rm{bg}} = 1.58~\rm{V}$ ($\nu_{\rm{B}} = 2$). The pajama for the inner edge channel is shown in Fig. \ref{figpajSimul1p5V}a, from which the oscillation frequency is derived and shown in Fig. \ref{figpajSimul1p5V}b. Fig. \ref{figpajSimul1p5V}c presents the pajama map for the outer channel, where the oscillation frequency primarily reflects electron pairing. This enables the extraction of the pairing frequency (Fig. \ref{figpajSimul1p5V}e) and the subtraction of the frequency associated with the bare outer channel (Fig. \ref{figpajSimul1p5V}d). Fig. \ref{figpajSimul1p5V}f shows the pajama map for two conductance oscillations with frequencies corresponding to the outer channel and pairing mode. The clear frequency beating observed here is similar to that seen in Fig. \ref{figpajSimul1p5V}c.

Similarly, Fig. \ref{Figure_S6} presents simulations of the pajama maps featuring electron tripling at $V_{\rm{bg}} = 2.5~\rm{V}$ ($\nu_{\rm{B}} = 2.9$). The calculated pajama for the inner channel, shown in Fig. \ref{Figure_S6}b, is derived from Fig. \ref{Figure_S6}a. Partitioning the middle edge channel reveals a pajama that predominantly exhibits oscillations of the middle channel, allowing for the extraction of the bare middle channel frequency (depicted in Fig. \ref{Figure_S6}c). The pairing frequency is determined by summing the frequencies of the inner and middle channels, as shown in Fig. \ref{Figure_S6}f, which matches the observed pajama in Fig. \ref{Figure_S6}c. The pajama map for the outer channel, presented in Fig. \ref{Figure_S6}g, is predominantly influenced by the tripling signal, enabling the extraction of the tripling frequency. With the frequencies of each edge channel known, Fig. \ref{Figure_S6}h shows the calculated pajama by combining the frequencies from tripling and the outer edge channel.

\section{Oscillation of the inner channel at $V_{\rm{bg}} =1.58~\rm{V}$ ($\nu_{\rm{B}}=2$) }
\label{secLobeStructure}

\begin{figure*}[ht]
	\centering
		\includegraphics[width=12cm]{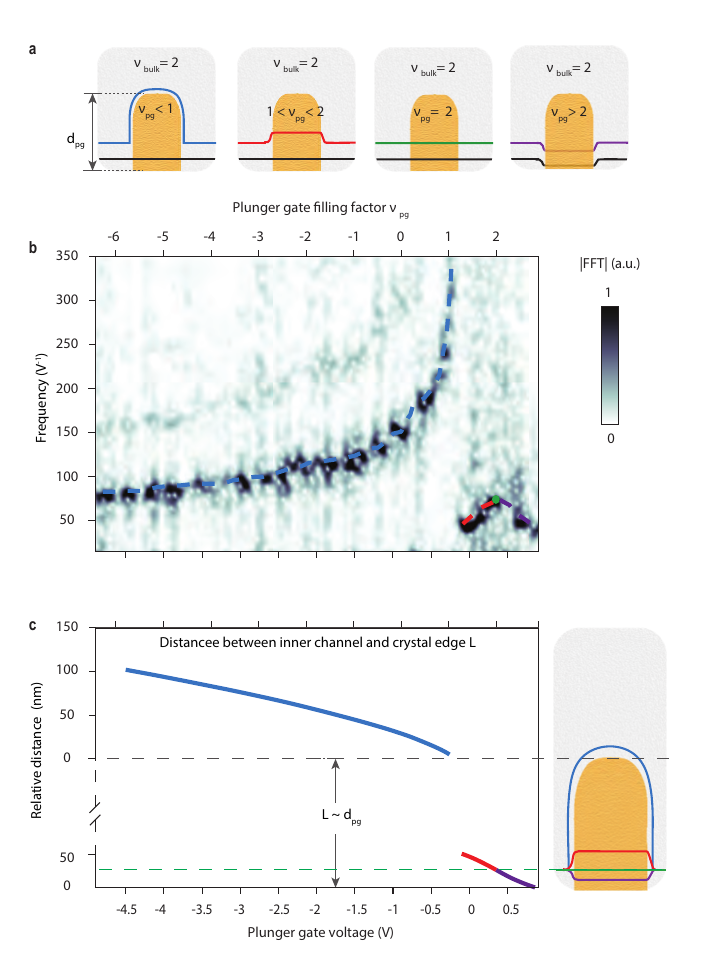}
	\caption{\textbf{Interference of the inner edge channel at $V_{\rm{bg}} = 1.58~\rm{V}$ ($\nu_{\rm{B}}=2$)}. \textbf{a}, Schematic diagrams illustrate four representative edge configurations. The solid black line represents the outer edge channel, while the colored lines indicate the inner edge channel at various plunger gate voltages. \textbf{b}, Plunge gate spectroscopy, extracted from the inner edge channel interference. \textbf{c}, Calculated displacement of the inner edge channel from the graphene crystal edge, derived from the data shown in panel \textbf{b}. The four colors correspond to the four inner edge configurations illustrated in panel \textbf{a}.}
	\label{fighighnupg}
\end{figure*}

Here, we describe the extra frequency line observed in Fig. 4c, which appears for the plunger gate filling factor between 1 and 2 (see top axis in Fig. 4c) and exhibit frequencies above 250 $\rm{V}^{-1}$. This contribution comes from the combination of multiple channels involved at this bulk filling factor of 3, resulting in a tripling mode.

To clarify this frequency line within the context of the plunger gate spectroscopy, we describe a simpler configuration at $\nu_{\rm{B}}=2$ with inner edge channel interference. The specific regime to be considered corresponds to the interfering edge channel remaining beneath the plunger gate. We show in Fig. \ref{fighighnupg}b that a frequency line around 50 $\rm{V}^{-1}$ emerges for $\nu_{\rm{pg}}>1$, before diverging at $\nu_{\rm{pg}}=1$ which corresponds to the expulsion of the edge channels from beneath the gate. This non-monotonic frequency dispersion is marked by the red and purple dashed lines. In this regime, the edge channel slightly moves beneath the gate upon changing the gate voltage, leading to changes in the area and resulting in low frequency conductance oscillation due to a small capacitive coupling.

The displacement of the edge channel relative to the crystal edge varies depending on the plunger gate voltage. For $V_{\rm{pg}}>0.4$ V ($\nu_{\rm{pg}}>2$), the edge channel moves closer to the crystal edge 
  due to the accumulation of electron with respect to bulk above this gate voltage value. Conversely, for $V_{\rm{pg}}<0.4$ V ($\nu_{\rm{pg}}<2$), the edge channel shifts inward toward the bulk as illustrated in Fig. S7a. This non-monotonous variation in dispersion with gate voltage reaches a maximum exactly at $\nu_{\rm{pg}}=2$, which corresponds to an iso-density condition both beneath the gate and within the bulk.

Fig. \ref{fighighnupg}c Shows the resulting integrated edge channel displacement as a function of the plunger gate voltage (see Methods), relative to the crystal edge (within a magnetic length \cite{coissard2023}). The maximum in frequency dispersion observed in Fig. S7b is reflected in Fig. S7c as an inflection point separating the red and purple segments. The spatial gap separating the red and the blue lines corresponds to the length of the plunger gate, over which the edge channel is expelled when $\nu_{\rm{pg}}$ approaches 1.

\section{Supplementary videos}
\label{secVideo}

\subsection{SI Video 1 (complement to Fig. 1g): Bias voltage evolution of the inner edge channel conductance oscillation}
\label{secVideoCond}
Conductance oscillation as a function of plunger gate voltage when partitioning the inner channel (as depicted in Fig. 1b) extracted for each voltage from the IV curves. 

\subsection{SI Video 2 (complement to Fig. 1h): Bias voltage evolution of the inner edge channel gate-spectroscopy}
\label{secVideoCondInner}

Bias voltage dependence of the Fourier transform (FT) of Fig. 1g versus plunger gate voltage when partitioning the inner edge channel. FT is computed with the data of SI Video 1. 

\subsection{SI Video 3 (complement to Fig. 2a): Bias voltage evolution of the outer edge channel gate-spectroscopy}
\label{secVideoCondOuter1}

Bias voltage dependence of the Fourier transform of Fig. 3a as a function of plunger gate voltage when partitioning the outer edge channel. 

\subsection{SI Video 4 (complement to Fig. 2c): Bias voltage evolution of the outer edge channel pajama map}
\label{secVideoCondOuter2}

The pajama map is extracted for each voltage from the IV curves, showcasing the partitioning of the outer channel (as demonstrated in Fig. 2c). The beating pattern aligns with the lobe structure of Fig. 2b (the lobe structure extracted from the pajama is almost identical). At high (positive or negative) voltage, a single frequency (tripling) dominates with a small modulation on top. Around $\pm 65~\upmu \rm{V}$, the two frequencies have similar amplitudes, and the pajama looks like a checkerboard. Around $- 30~\upmu \rm{V}$, the pairing signal dominates, and the pajama is similar to the one at high (positive or negative) voltage. Around $0~\upmu \rm{V}$, the pajama pattern is akin to the one in Fig. 2c, with one dominating oscillation (pairing) but periodically interrupted by a second smaller oscillation stemming from the outer edge channel.

\subsection{SI Video 5 (complement to Fig. 4c): Bias voltage evolution of the outer edge channel (electron tripling) gate-spectroscopy}
\label{secVideoCondTripling}
Conductance oscillations with plunger gate voltage, obtained by partitioning the outer edge channel at filling factor 3 (Fig. 4c), are extracted for each voltage from the IV curves. At high (positive or negative) voltage, a single frequency (tripling) is visible. Around $\pm 65~\upmu \rm{V}$, a second lower frequency appears, corresponding to the pairing of the middle and outer channels. Around $\pm 30~\upmu \rm{V}$, two additional frequencies appear, corresponding to the middle and outer edge channel oscillations. Around $\pm 20~\upmu \rm{V}$, all four amplitudes are fainter, corresponding to the first minimum of the lobe structure. Around $\pm 0~\upmu \rm{V}$, one recovers Fig. 4c. Similar to Supplementary Video 3, all three signals appear and disappear with comparable frequency (in bias voltage), in agreement with the lobe structure of Fig. \ref{figLobeTripl}a. Note that the different frequency dispersions are independent of bias voltage.

\subsection{SI Video 6 (complement to Fig. 4d): Bias voltage evolution of the middle edge channel (electron tripling) gate-spectroscopy}
\label{secVideoCondMiddle}
Conductance oscillations with plunger gate voltage, obtained by partitioning the middle edge channel at filling factor 3 (as in Fig. 4d), are extracted for each voltage from the IV curves. At high (positive or negative) voltage, two signals of comparable amplitude are visible, corresponding to middle and inner edge channel oscillations. As the voltage is increased, both two amplitudes increase, while a new pairing signal appears around $\pm 55~\upmu \rm{V}$. Around $\pm 0~\upmu \rm{V}$, one recovers the map of Fig. 4d. Similar to supplementary videos 3 and 5, middle and inner signals appear and disappear with comparable frequency (in bias voltage), in agreement with the lobe structure of Fig. \ref{figLobeTripl}b. Note that the different frequency dispersion is independent of bias voltage.

\subsection{SI video 7: Bias voltage evolution of the outer edge channel pajama at $V_{\rm{bg}} = 2.6~\rm{V}$}
\label{secVideoPajInner1}
Conductance oscillations with plunger gate voltage and magnetic field, obtained by partitioning the inner edge channel at filling factor 3.1 (as in Fig. S5d), are extracted for each voltage from the IV curve. At high (positive or negative) voltage, the interference pattern exhibits a negative slope in the $B$-$V_{\rm{pg}}$ plane, characteristic of Coulomb-dominated interference due to a contribution from the inner edge channel. Between $\pm 80~\upmu \rm{V}$, the outer channel, pairing (middle and outer), and tripling can contribute to the interference. The pajama pattern thus changes with bias voltage depending on which channel is dominating.

\subsection{SI video 8 (complement to Fig. S4d): Bias voltage evolution of the inner edge channel pajama at $V_{\rm{bg}} = 2.33~\rm{V}$}
\label{secVideoPajInner2}
The pajama map, obtained by partitioning the inner edge channel at filling factor 3 (as in Fig. S4d), is extracted for each voltage from the IV curve. The interference pattern exhibits a negative slope in the $B$-$V_{\rm{pg}}$ plane, characteristic of Coulomb-dominated interference. The interference pattern is nearly independent of bias voltage. The slight modifications in the pattern with bias voltage are primarily attributed to changes in visibility.

\subsection{SI video 9 (complement to Fig. S4e): Bias voltage evolution of the middle edge channel pajama at $V_{\rm{bg}} = 2.33~\rm{V}$}
\label{secVideoPajMid2}
The pajama map, obtained by partitioning the middle edge channel at filling factor 3 (as in Fig. S4e), is extracted for each voltage from the IV curve. As illustrated in Fig. S4e, interference from the inner and the middle channel, and their pairing can occur in this scenario. The pajama pattern thus varies with bias voltage, depending on which channel is dominating at a given voltage. At high (positive or negative) voltage, the middle edge channel dominates, giving rise to an AB-dominated pajama. As the voltage increases, the inner channel appears (Coulomb dominated) and introduces an extra modulation around $\pm 110~\upmu \rm{V}$. Around $\pm 77~\upmu \rm{V}$, the pairing channel appears, dominating the overall low-visibility interference pattern. Around $\pm 50~\upmu \rm{V}$, the middle channel dominates the oscillations with a small modulation due to pairing. Around $0~\upmu \rm{V}$, all channels are present with low visibility.

\subsection{SI video 10 (complement to Fig. S4f): Bias voltage evolution of the outer edge channel pajama at $V_{\rm{bg}} = 2.33~\rm{V}$}
\label{secVideoPajMid1}
Conductance oscillations with plunger gate voltage and magnetic field, obtained by partitioning the middle edge channel at filling factor 3 (as in Fig. S4f), are extracted for each voltage from the IV curve. As illustrated in Fig. S4f, interference from the outer channel, the pairing (involving the middle and outer channels), and the tripling can occur in this scenario. The pajama pattern thus varies with bias voltage, depending on which channel is dominating at a given voltage.

\clearpage

\end{document}